**Article type:** **Original Article**

# Tunable Conformal Graphene Growth on Oxide Nanotube scaffolds: Towards Superwettable Hierarchical 2D–3D Architectures

Fernando Núñez-Gálvez,*[1,2] Muhammad Hamza,[3] Johannes Berndt,[3] Thomas Strunskus,[4] Franziska Traeger,[5] Nicoló di Novo,[2] Juan R. Sánchez-Valencia,[2] Carmen Lopez-Santos,[1,2] Ana Borras,*[2] and Eva Kovacevic[3]*

[1] Departamento de Física Aplicada I, Escuela Politécnica Superior, Universidad de Sevilla (US)

[2] Nanotechnology on Surfaces and Plasma Lab, Materials Science Institute of Seville (Spanish National Research Council (CSIC) – Universidad de Sevilla (US))

[3] GREMI, Groupe de Recherches sur l'Energétique des Milieux Ionisés (CNRS-Université d'Orléans)

[4] Department of Materials Science, Kiel University, Kaiserstr, 2, 24143 Kiel, Germany

[5] Westfälische Hochschule, August-Schmidt-Ring 10, 45665 Recklinghausen, Germany

*Corresponding author. E-mail : eva.kovacevic@univ-orleans.fr

## Highlights:

- Dry process production of tunable 2D–3D architectures by conformal vertical graphene growth on diverse oxide nanotube scaffolds.
- Hierarchical 2D–3D architectures combine conductivity with high surface accessibility.
- Re-entrant structures deliver durable superomniphobicity.

## TOC

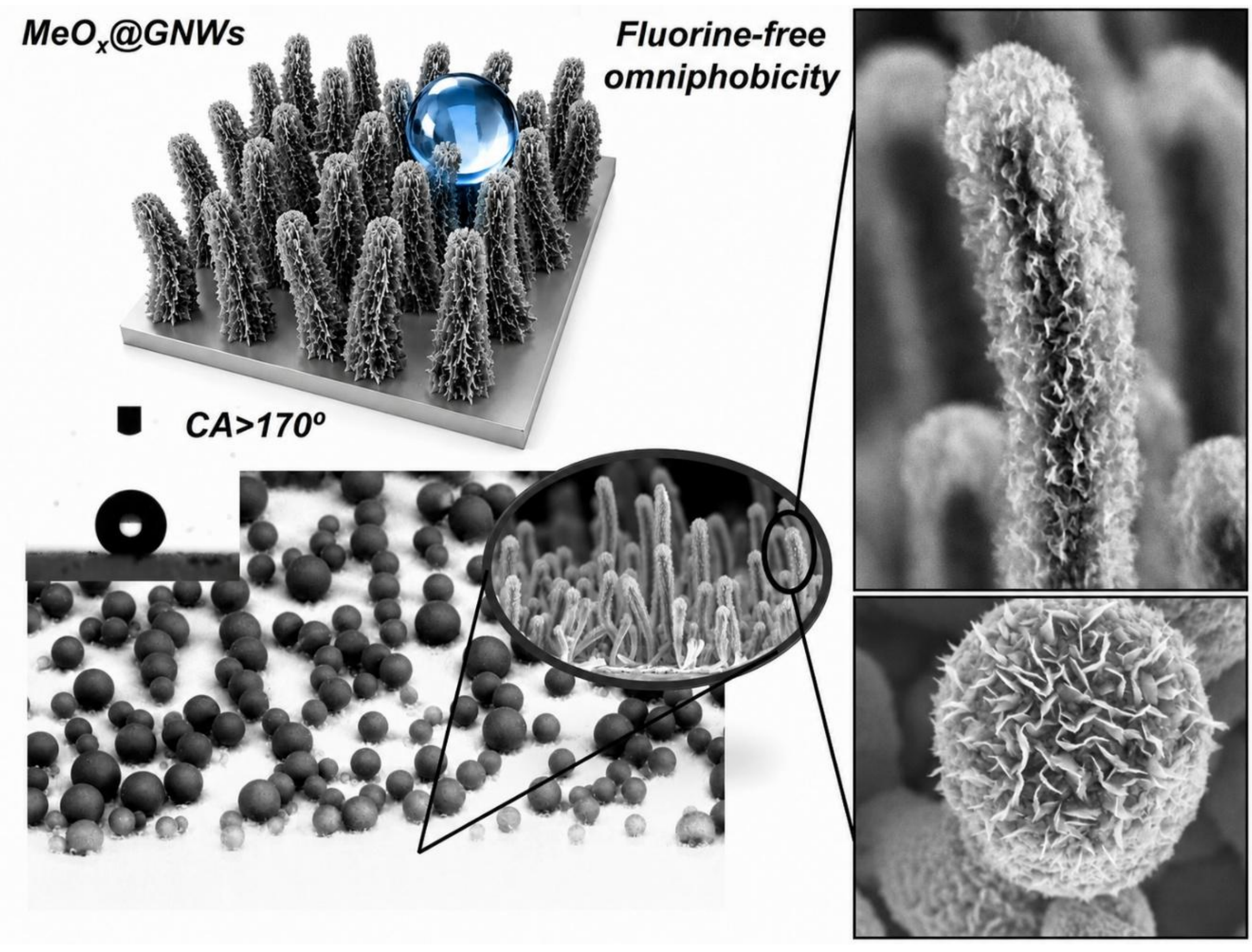

## Abstract

Hierarchical hybrid nanoarchitectures that integrate vertically oriented graphene nanowalls (GNWs) with metal oxide ($MeO_x$) nanotube scaffolds offer versatile platform for smart surfaces, nanoelectronics, and electrochemical technologies. Herein we present rapid, dry, plasma-assisted fabrication route that enables direct and conformal growth of GNWs on mechanically robust $MeO_x$ nanoforests. The method combines supported single-crystalline organic nanowires as a 1D soft template with sequential plasma-enabled oxide deposition and GNW growth, all performed under mild temperature, power, and vacuum conditions. This approach yields an unprecedented 2D–3D hierarchical architecture consisting of tunable-thickness $MeO_x$ nanotubes uniformly decorated with radially oriented graphene nanosheets, forming re-entrant, multiscale surface. Resulting hierarchical roughness imparts fluorine-free, long-term omniphobicity, with contact angles exceeding 170° for water, bovine serum, and other complex fluids. GNWs dominate the wetting response across $TiO_2$, $Al_2O_3$, and $SiO_2$ nanotube scaffolds, effectively decoupling surface behavior from intrinsic oxide chemistry and maintaining robust repellency under UV irradiation and water condensation. Comprehensive SEM, TEM, XPS, angle-resolved NEXAFS, and Raman analyses elucidate growth mechanism and confirm preservation of the $sp^2$ graphitic framework, together with controlled degree of edge functionalization. Overall, this work establishes universal, substrate-compatible, low-temperature, and scalable route for the fabrication of tunable graphene-metal oxide nano/microstructured multifunctional surfaces.

# 1 Introduction

The growth of vertically oriented graphene nanowalls (GNWs) on nanostructured substrates has gained significant attention due to the appealing intrinsic properties of each component and the anticipated synergistic effects arising from their combination, such as enhanced electric-field effects, increased surface activity, and improved structural functionality. However, although the plasma-assisted synthesis of vertical graphene on flat substrates, carbon nanofibers, and nanoparticles is well established, [1],[2],[3],[4],[5] the integration of graphene with metal-oxide ($MeO_x$) nanotube architectures remains largely unexplored. Only a few studies, primarily focused on ZnO nanotubes for sensing applications,[6],[7] have demonstrated such hybrid systems,  evidencing the absence of reliable synthetic routes to produce these hierarchical nanoarchitectures.

On one hand, vertically oriented graphene consists of few-layer graphene sheets vertically aligned (with controllable height), anchored to the substrate surface, resulting in a high aspect ratio morphology that offers a large accessible surface area, naturally exhibits a high density of exposed edges, excellent electronic conductivity, excellent electron transport pathways, and chemical reactivity at the exposed edges.[8], [9], [10] These unique characteristics make GNWs highly attractive for a broad range of applications, including supercapacitors,[11],[12] batteries,[13], [14] field emission devices,[15] nanogenerators,[16] biosensors,[17] gas sensors,[6] and electrocatalysis.[18] However, the ability to synthesize high-quality GNWs at low temperatures without catalysts remains a critical challenge for advancing their integration into flexible electronics, CMOS-compatible processes, and other sensitive platforms.[19], [20]

On the other hand, metal oxides 1D nanostructures have become indispensable in advanced materials science due to their transport and optical properties and enhanced interfacial phenomena, that emerge at the nano scale. The controlled fabrication via plasma-assisted vacuum techniques and soft-templates[21] enables precise morphological control in $MeO_x$ nanotubes and core@shell nanostructures, which is critical for tailoring not only surface interactions but also functional responses, thanks to the under-design nature of the shells. Highly porous nanostructured networks can produce surfaces of directed pathways for light and charge transport, yielding high-performance sensors,[22] photonics[23] and energy systems,[24] as well as surfaces with controlled wettability[25] relevant to energy, self-cleaning, and microfluidics applications.[26]

Particularly, hierarchical 2D/3D nanostructures, such as nanotube arrays, nanopillars, and porous frameworks, are particularly effective in stabilizing Cassie Baxter wetting by reducing the effective solid–liquid contact fraction and promoting air entrapment.[27], [28] Unfortunately, for intrinsically high surface energy materials such as metal oxides, the Cassie–Baxter state is often metastable and prone to collapse into a Wenzel regime under real environmental conditions [29], [30] or the action of external stimuli such as irradiation, condensation, or chemical exposure.[25], [31] To address this limitation, two-dimensional (2D) carbon nanomaterials, most notably graphene, have emerged as promising candidates for surface functionalization due to their low surface energy, chemical inertness, and mechanical robustness.[32], [33], [34], [35]

Among super wettable systems, i.e., materials designed to exhibit extreme wetting behaviours,[36], [37] omniphobic surfaces capable of repelling not only water but also complex, low-surface-tension fluids, pollutant particles or biological agents remain challenging to realize.[38] Traditional approaches rely on fluorinated coatings or organic low-surface-energy molecules.[39] Fluorinated compounds, however, pose significant environmental risks, and tightening regulations on per- and polyfluoroalkyl substances are expected to further restrict their use.[40] Organic coatings, in turn, often lack long-term stability, showing limited durability and susceptibility to chemical degradation. As a result, there is growing interest in fluorine-free approaches that combine engineered surface architectures with robust chemistries to achieve durable, environmentally benign omniphobicity.[41]

Beyond wetting control, such hybrid 2D/3D architectures open new opportunities in water harvesting and water-energy nexus applications,[42] particularly in the emergent topics of triboelectric[43] and moisture energy generators.[44] Hierarchical nanostructures that regulate condensation, droplet mobility, and liquid infiltration can significantly enhance interfacial transport phenomena, while graphene-based materials provide electrically conductive pathways and chemical stability. Thus, surfaces that simultaneously exhibit tunable wetting behavior and robust environmental resilience are highly attractive for next-generation moisture-responsive and energy-harvesting devices. However, as mentioned above, the integration of graphene nanowalls (GNWs) with high-aspect-ratio 3D oxide nanotube architectures remains scarcely explored. In particular, the synergistic combination of 2D graphene nanowalls conformally decorating 3D hierarchical nanotube networks represent a largely unexplored materials platform. Within this scope, the multistep plasma-assisted growth offers unique advantages. Unlike transferred graphene sheets, polymer-assisted coatings or CVD methods,[45], [46], [47] plasma-enhanced chemical vapor deposition (PECVD) has emerged as a powerful and versatile method for the growth of GNWs [2], [48]. This fabrication process, relying on solventless, mild-temperature plasma-based growth of graphene, renders these materials compatible with a broad spectrum of substrates, including patterned, flexible, and thermally fragile substrates. In particular, the use of capacitively coupled radiofrequency (RF) plasma systems enables precise control of plasma parameters such as energy density, ion bombardment, and gas-phase composition, all of which strongly influence the nucleation, orientation, and structural quality of GNWs, which opens the path to fabricate them on $MeO_x$ supported nanotubes.

In this work, we take a step forward in the exploitation of GNWs and present a detailed experimental study demonstrating control over the fabrication of graphene-decorated hierarchical surfaces from 1D templates. This leads to a fluorine-free strategy for fabricating superomniphobic surfaces by integrating graphene nanowalls with oxide nanotube arrays ($TiO_2$, $SiO_2$, and $Al_2O_3$) via a controlled multistep plasma and vacuum deposition process. The resulting $MeO_x$ NTs@GNWs surfaces exhibit pronounced hierarchical roughness, re-entrant geometry, and reduced surface energy, enabling a stable Cassie–Baxter wetting state across a wide range of liquids, including water, biological fluids, and organic solutions. We systematically investigate the wetting stability under UV–IR irradiation, condensation at subzero temperatures, and exposure to complex liquids, revealing the critical role of graphene nanowalls in suppressing Cassie–Baxter to Wenzel transitions.[49] Furthermore, the surfaces demonstrate outstanding repellent behavior and resistance to droplet pinning, highlighting their potential for applications in environmental protection, condensation management, and moisture-driven energy systems.

# 2 Experimental Section

## 2.1. 3D Nanostructures synthesis

Fabrication of the 3D metal-oxides nanostructures comprised a full-vacuum multi-step procedure which can be conveniently divided into three main steps as described in refs [21], [23], [85]. *Seeds deposition*: $TiO_2$, $SiO_2$ and $Al_2O_3$ between 200-300 nm thin films were fabricated by Plasma Enhanced Chemical Vapour Deposition (PECVD) at mild temperature. Titanium tetraisopropoxide (TTIP), methyltrichlorosilane ($Cl_3MeSi$) and trimethylaluminium (TMA) were utilized as precursors (Sigma Aldrich). TTIP, $Cl_3MeSi$ and TMA were used as Ti, Si and Al precursors respectively. Those precursors were dosed to the reactor through a regulable valve, and a shower-like dispenser positioned over the samples. TTIP and TMA and were kept at 40ºC during the deposition while low vapor pressure $Cl_3MeSi$ precursor was kept at RT during deposit. Tubes and lines were heated up to 80ºC and also the shower. 20 sccm of oxygen was used as plasma gas and inserted in the reactor by using a mass flow controller. Base pressure of the chamber was under $10^{-6}$ mbar and the total pressure during deposition was between 1.5-2 x$10^{-2}$ mbar. So, the partial pressure of the precursor during deposition was around 1.3x10-3 mbar. The plasma was generated in a 2.45 GHz microwave Electron-Cyclotron Resonance (ECR) SLAN-II source operating at 400 W and the deposition was carried out in the downstream region of the reactor. The growth rate was settled at 2,3-3 nm/min for $SiO_2$ and $TiO_2$ and 10 nm/min for $Al_2O_3$. The thicknesses of all the thin films were controlled by simply adjusting the deposition time to get the desired thickness after calibration by SEM on a Si (100) reference substrate**.** *Organic scaffold*: The PVD procedure for the formation of single crystal ONWs has been fully described in previous references. Organic NWs were grown by vacuum deposition of $H_2$-Phthalocyanine powder (Sigma-Aldrich) using an LTE01 evaporation source from Kurt J. Lesker from a Knudsen cell. The base pressure of the chamber was lower than $10^{-6}$ mbar. The total pressure during deposition was 3x$10^{-3}$ mbar of Ar, the deposition rate was 0.45 Å/s (measured by QCM) and the substrate temperature was 200-210 °C. The sample-to-evaporator distance was 6.5 cm. The nominal thickness of 1.00 kÅ which corresponds to 2-5 μm of nanowire length. *Shell fabrication*: Shells were fabricated in a same way described in seeds deposition. *Empty of the 1D nanostructures*. As in subsequent processes, the deposition of graphene nanosheets, substrate temperatures above 600°C is reached, this temperature is more than sufficient to completely remove the organic inner filler of the nanotube, i.e. phthalocyanine. This is removed through the intrinsic porosity of the material itself so that the core remains empty forming the nanotubes.

## 2.2. GNWs deposition

The growth of vertical graphene or graphene nanowalls was carried out using a radiofrequency (RF) plasma-enhanced chemical vapor deposition (PECVD) system operating at 13.56 MHz in a capacitively coupled (CCP) configuration reported previously in references [53], [54] . The system consists of a cylindrical vacuum chamber equipped with two parallel stainless-steel electrodes, each with a diameter of 8 cm and a variable inter-electrode distance of 1.5 to 7 cm. This growth has been performed while sustaining the plasma at a low RF power of 30W. Samples were mounted on the grounded electrode, which is also heated up to 600° C for the growth. For growth, a base pressure in the range of ~10-7 mbar was achieved using a rotary pump, which was followed by a turbo molecular pump. During the growth process, working pressure was maintained between 0.2-0.5 mbar. To ignite the plasma, gases were introduced in the reactor through highly accurate mass flow controllers (model). Three gases were used in the whole process: Ethylene ($C_2H_4$) served as the carbon source, argon (Ar) as an inert carrier gas to dilute the $C_2H_4$, and hydrogen ($H_2$) to etch the amorphous carbon while promoting vertical growth. The growth process was completed in two steps, which is extensively explained in [53], [54]. The first step is to deposit the carbon on the walls of the reactor, referred to as the seasoning step, and in the second step, Ar plasma was ignited to etch the carbon from the walls of the reactor and deposit it on the substrate. To grow the GNWs, samples were mounted just before the second step.

## 2.3. Characterization techniques

SEM micrographs were acquired both in a Hitachi S4800 and in a Supra 40 Zeiss system. All samples were characterized without additional conductive coatings. Environmental Scanning Electron Microscope (ESEM) experiments were performed in Zeiss EVO by placed the sample on a Peltier stage and controlling the water vapor pressure in the system. Temperature was stabilized to -1ºC or -5ºC for condensation in the superhydrophobic samples. The samples were dispersed onto holey carbon films on Cu grids from Agar Scientific for TEM characterization. HRTEM was carried out with a FEI Tecnai G2F30 S-Twin STEM microscope working at 200 kV. Bright-field TEM images were obtained in a CM200 apparatus from Philips. HAADF-STEM images were acquired in a Tecnai G2F30 S-Twin STEM from FEI.

The structure and organization of graphene nanowalls were characterized by means of a Raman spectrometer (Renishaw Virsa, Standard Probe-VRP11-532) in ex-situ mode with a 532nm laser and a spot size of ~1.6 μm. XPS experiments were performed in a Phoibos 100 DLD X-ray spectrometer from SPECS. The spectra were collected using a nonmonochromatic Al Kα source for general survey spectra and high-resolution spectra. C 1s signal at 284.5 eV was utilized for calibration of the binding energy in the spectra. The assignment of the BE to the different elements in the spectra corresponds to the data in Refs. [74], [75] Glazing-Incident X-Ray Diffraction (GIXRD) was used in a Panalytical X'PERT PRO system for surface analysis.

Surface composition, chemical coordination, and orientation of the structures were investigated using angle-resolved near-edge X-ray absorption fine structure (NEXAFS) spectroscopy at X-ray incidence angles of 30°, 55°, and 90° relative to the substrate normal. The measurements were performed at the HE-SGM monochromator dipole magnet beamline at the BESSY II synchrotron radiation source (Berlin, Germany). NEXAFS spectra were acquired at the C K-edge in the partial energy electron yield (PEY) mode with a retarding potential of -150 V. The raw spectra were divided by the monochromator transmission function obtained from a freshly sputtered Au sample. The monochromator energy scale was calibrated using the carbon K-edge $\pi^*$ transition of graphite, located at 285.38 eV. The spectra were normalized to the intensity value at an energy of around 292.12 eV.

Superwettable and superomniphobicity behavior of GNWs-coated nanostructures as well pristine references were tested using a goniometer OCA25 from DataPhysics with variable droplet volume (generally 2µl for contact angle measurements of different liquids). UV photoactivation test was carried out using a ASB-XE-175EX with a lamp type CERMAX LX175BF lamp, with a 5600 K color temperature and a power maximum of 200 W irradiating the samples disposed vertically faced to the light with a distance of 30 cm.

## 3 Results and Discussion

### 3.1 Step-by-step fabrication of hierarchical $MeO_x$@GNWs by plasma supported deposition.

In this article, we have employed a soft-template method using supported single-crystalline organic nanowires (ONWs) to fabricate hierarchical $MeO_x$@GNWs nanotubes. The use of ONWs as vacuum-processable, solventless, removable templates has been exploited over the last few years in combination with the conformal growth of functional layers by plasma-supported deposition methods for the formation of core@shell and core@multishell nanowires and nanotubes, with applications in solar cells,[50] nanogenerators,[24] sensors,[22] and superhydrophobic surfaces.[26] Herein, we go a step further and demonstrate the advantages of the method for the development of three-dimensional nanoarchitectures, exploiting the surface of $MeO_x$ nanotubes as nucleation sites for the formation of vertical graphene nanowalls grown under mild plasma and vacuum conditions. **Scheme 1** showcases the step-by-step schematic of the procedure.

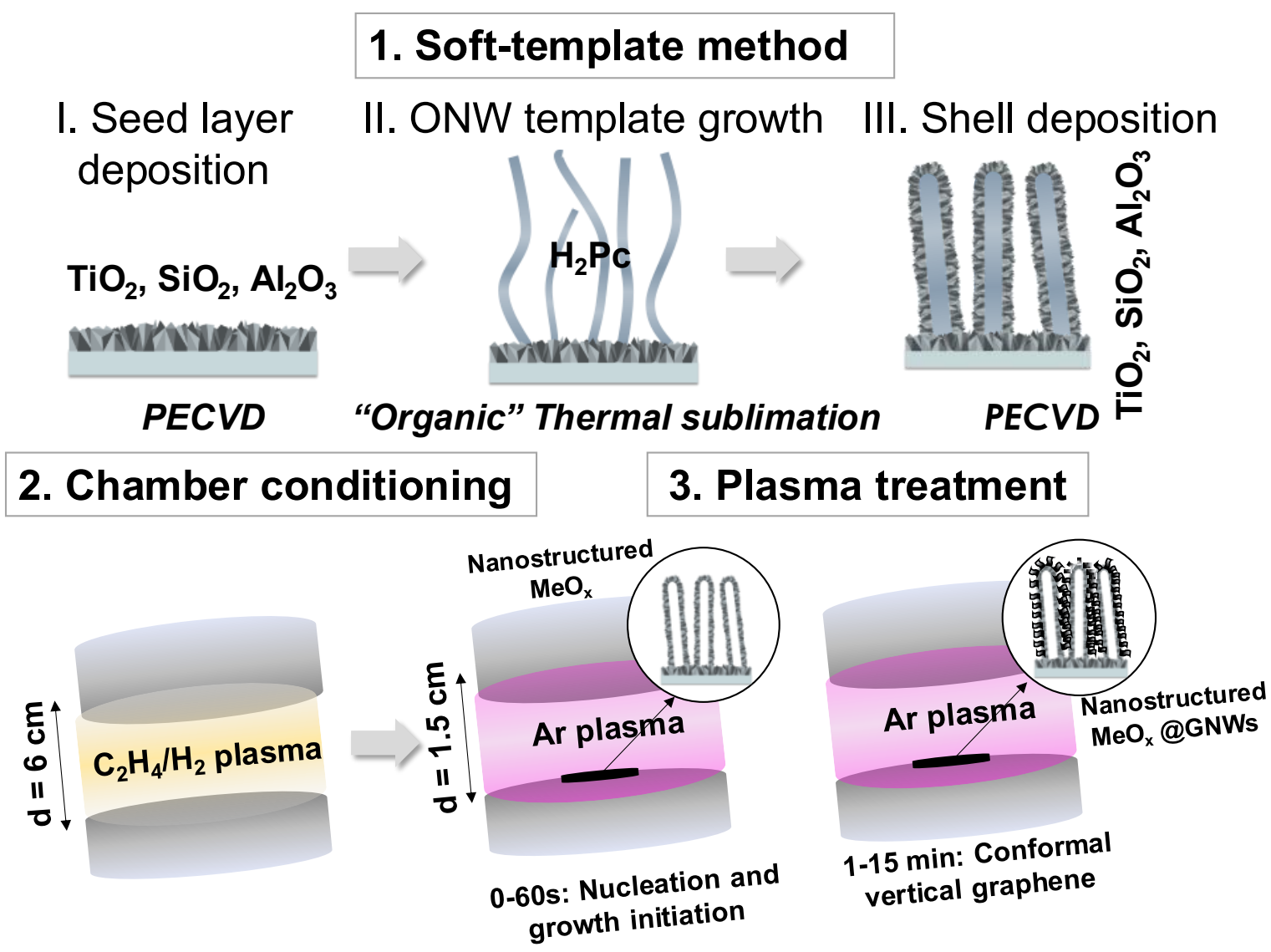

**Scheme 1.** Step-by-step formation of $MeO_x$ @GNWs enabled by vacuum and plasma processes.

Step 1: soft-template method for the formation of $MeO_x$ nanotubes, including seed film and shell deposition by PECVD (I and III), and growth of single-crystalline organic nanowires as a 1D template by thermal sublimation (II). Step 2: chamber conditioning for the precursor inject. Step 3: plasma assisted vertical graphene growth. All the steps are carried out under mild temperature conditions (< 600 ºC), mild vacuum (<0.5 mbar) and plasma power (<30 W).

Step I represents the formation of a seed layer that offers nucleation sites for the growth of ONWs. The chemical nature of the seed layer is not restricted, as the growth mechanism is purely driven by crystallization that couples to the directionality of π-stacking among extended aromatic molecules. Hence, the formation of ONWs from molecules such as porphyrins, phthalocyanines, and perylenes has been reported on diverse substrates and seed layers, [51] including $MeO_x$ and transparent conducting oxide (TCO) thin films, metal nanoparticles, plasma-activated organic layers deposited on glass, Si, and metal alloys, as well as cellulose and polymers. Herein, for the sake of simplicity and as a first demonstration of the potential of the approach, we have employed as seed layers the same plasma-enable $MeO_x$ forming the shell of the nanowires. **Figure S1** shows characteristic cross-sectional SEM micrographs of these $MeO_x$ layers, including $TiO_2$ and $SiO_2$ compositions (see Experimental Section for synthesis details). The thickness of the seed layers has been kept below 250 nm, although this parameter is not restrictive, as the surface roughness provides sufficient nucleation sites for the ONWs (a major parameter controlling the final density). The $TiO_2$ and $SiO_2$ layers are formed by nanoscale columnar stacks, developing a cauliflower-like surface with RMS in the range of a few nanometers[52]. The second step (Scheme 1, Step II) involves the low-thermal evaporation of the small-molecule precursors to form the ONWs, which can be performed under high or mild vacuum in an inert atmosphere. During this second step, the temperature of the substrates is kept between 0.5 and 0.8 times the evaporation temperature of the organic molecules. These conditions ensure the formation of flexible single-crystalline nanowires with lengths, thicknesses, and densities controlled by the substrate temperature, pressure, deposition rate, duration, surface tension, and roughness. A versatile molecule for the formation of these ONWs is the macrocyclic, metal-free phthalocyanine ($H_2Pc$), which is low-cost, easily evaporated at temperatures below 380 ºC (i.e., compatible with ONW formation at temperatures below 220 ºC) and readily removable to form nanotubes by mild annealing in air or vacuum as we will explain below. **Figure S2** shows the statistical distribution analysis for the as-grown nanowires on the three selected seed layers for lengths ((2.6 ± 0.6) µm, (4.6 ± 1.1) µm and (3.3 ± 1.1) µm), diameters ((240 ± 35) nm, (565 ± 100) nm and (265 ± 65) nm ) and densities ((1.6 ± 0.2) NW/$\mu m^2$, (0.9 ± 0.1) NWs/$\mu m^2$ and (4.3 ± 0.9) NWs/$\mu m^2$) for $SiO_2$ NTs, $TiO_2$ NTs and $Al_2O_3$ NTs. The as-grown nanowires are randomly oriented, consistent with their flexible nature. However, they tend to form a vertical array during Step III, as a consequence of the orientation driven by the electric field in the plasma sheath formed at the interface with the substrate during the conformal growth of the $MeO_x$ layer by plasma enhanced chemical vapor deposition (see **Figure 1** a), d), and h)). The surface morphology of the seeds is replicated on the conformal layers (shells), but on a smaller scale. The surface of the core@shell nanowires exhibits a globular microstructure with features on the order of 20-30 nm across the three architectures.

In Steps 2 and 3, graphene nanowalls are grown on the surfaces of core@shell nanowires via low-pressure PECVD. [53],[54],[55], ,[56],[57], [58] The method provides well-conductive, well-organized vertically aligned GNWs with substrate temperature already after 450°C using low-power plasma conditions (<30 W). This flexible process does not require a specific substrate geometry or a catalyst.[55] The majority of neutral gases remain at room temperature and are not excited. The growth mechanism was attributed to the delicate balance between precursor introduction, either by sputtering or by plasma-volume synthesis of hydrocarbon species, together with the wall-mediated chemistry that governs precursor recycling and radical quenching (Scheme 1, Step 2 Chamber Conditioning). This interplay is essential for stabilizing the plasma environment, ensuring a consistent supply of reactive carbon fragments, and ultimately enabling reproducible and controllable nanostructure growth.[54] For each set of parameters, growth was performed simultaneously on flat reference substrates (see cross-sectional and top views of flat silicon in **Figure S3**) and on the oxide nanotube, under identical conditions to ensure better control of the material, comparison, and reproducibility analysis. The growth height and density can be controlled by plasma-on time and surface temperature, respectively. For the parameters settled herein (GNWs grown at 600ºC in standard Ar plasma after a $C_2H_2/H_2$ wall treatment for 15 min of plasma-on time), the height of well-aligned and interconnected vertical nanostructures is ca. 220-240 nm (Figure S3).

This formation is also extended to the nanostructured substrates. Thus, Figure 1 compares the nanostructures before (panels a), d) and h)) and after the formation of the GNW for $SiO_2$ NTs@GNWs (b), $TiO_2$ NTs@GNWs (e), and $Al_2O_3$ NTs@GNWs (i), respectively. The GNW growth conforms to the nanotube diameter and is radial along it, i.e., producing a 2D-on-3D nanoarchitecture, without altering their vertically standing morphology.

It is well known that the formation of GNWs does not follow a linear growth rate because both growth and etching processes begin to compete at a certain point. [53], [54] It is also important to mention that self-shadowing from the vertical alignment of the nanotubes can produce an uneven growth of the graphene shells along the length of the tubes, with higher thickness at the top. No self-shadowing effect is observed, and a full coverage of graphene nanowalls is achieved on nanowires that remain aligned with the plasma direction. This alignment enables graphene growth along the entire length of the nanowire, which is particularly evident for $SiO_2$ (Figure 1 b-c) or $Al_2O_3$ (Figure 1 i-j) nanotubes, whereas it is less pronounced for $TiO_2$ nanotubes (Figure 1 e-f). In the case of $SiO_2$ nanotubes, graphene nanowalls grow along the entire length of the nanowires, including the bottom region and interface with the substrate, indicating that plasma species can effectively penetrate the nanotube array when no self-shadowing is observed (**Figure S4**). Figure 1 c), f) and j) and **Figure S5** present detailed single nanotubes after the growth of GNWs. The preferential direction of growth can be observed, from the core to the outer part, following the surface nano-roughness of the different shell materials. Finally, Figure 1 k) shows a top-view of a highly vertically aligned $Al_2O_3$ NTs@GNWs nanoforest. Comparing the different nanotube compositions, it can be concluded that while the morphology of GNWs is similar and independent of the material nanostructure and chemical composition, their conformality strongly depends on the type of $MeO_x$ NTs, and consisting with a shadowing-mediated growth mechanism. A full elucidation of this effect requires a dedicated analysis of the growth conditions, as aimed at in the next section.

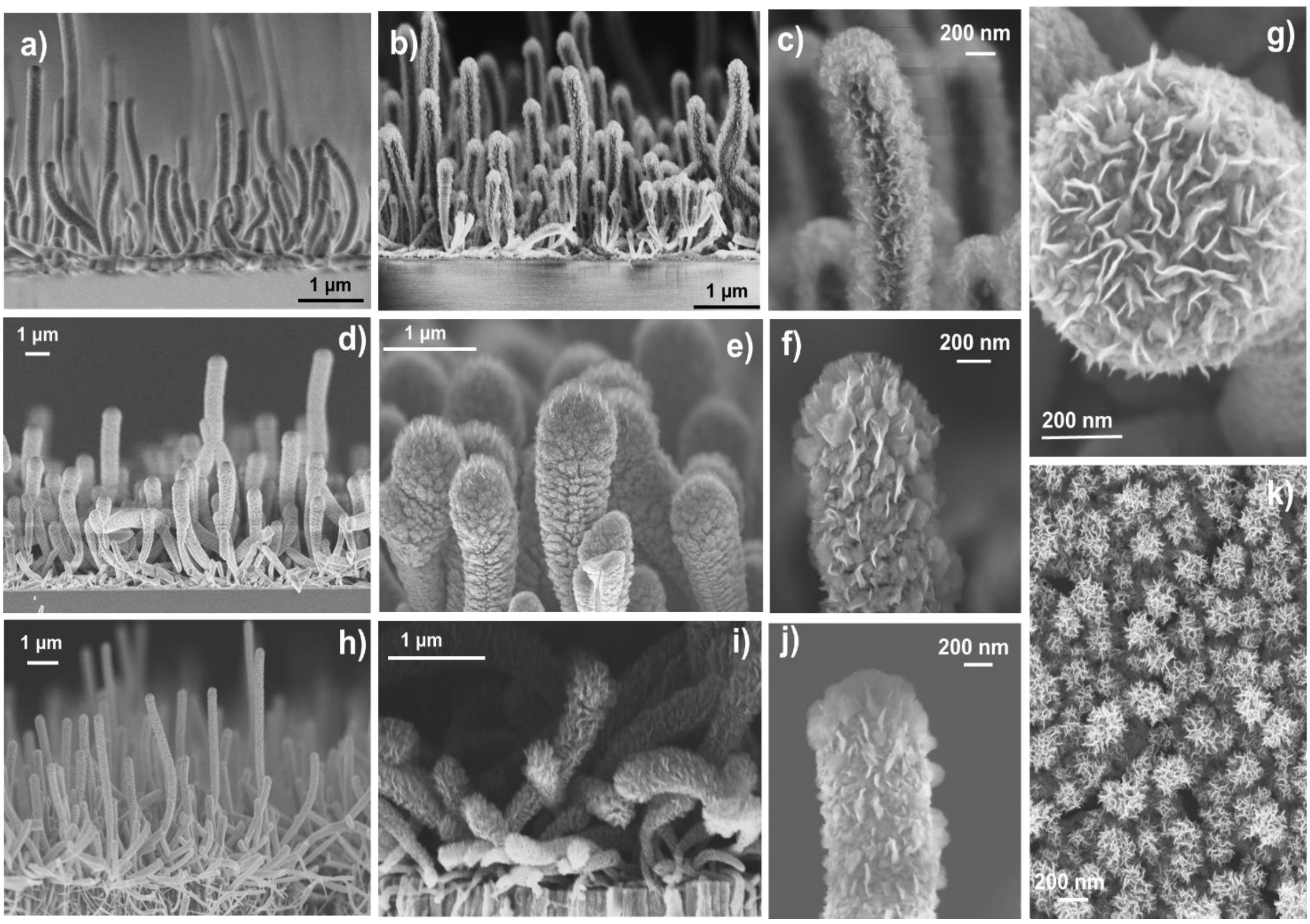

**Figure 1. Microstructure of the 3D Hierarchical $MeO_x$@GNWs supported nanostructures.** Characteristic SEM micrographs showing the formation of GNWs on supported $MeO_x$ nanotubes on Si wafers for: a) as-grown $SiO_2$ nanotubes, b-c) $SiO_2$ NTs@GNWs (i.e., after deposition of GNW), d) $TiO_2$ NTs, e-f) $TiO_2$ NTs@GNWs, g) detailed top-view of a graphene-coated single $TiO_2$ nanotube, h) $Al_2O_3$ NTs before GNWs deposition, i-j) $Al_2O_3$ NTs@GNWs and k) top-view of the previous sample.

### 3.2 Advanced characterization and insights into the growth mechanism.

Substrate temperature plays a critical role in determining both the nucleation density and surface coverage of GNWs during the PECVD process on flat 2D surfaces. At higher temperatures of synthesis (600 °C, see Figure 1), enhanced surface diffusion of carbon adatoms promotes uniform nucleation and lateral continuity of the initial graphitic layer, which is a prerequisite for the subsequent transition to vertical nanowall growth. Increased thermal energy also facilitates the rearrangement of $sp^2$ carbon into energetically favourable configurations, leading to improved graphitization and sustained vertical growth across the entire three-dimensional surface. In contrast, there is also a critical limit of the GNWs formation of the 3D structures when the deposition temperature is below to 450 °C for the presented plasma parameters, the limited surface mobility of carbon species hinders the formation of a continuous initial graphene layer (see **Figure S7** for detailed SEM images). Consequently, GNWs nucleation becomes localized and discontinuous, resulting in reduced surface coverage and the presence of partially coated or uncoated regions. Moreover, insufficient thermal energy at lower temperatures suppresses the vertical growth transition, favouring the formation of amorphous or horizontally oriented carbon domains rather than well-defined nanowalls. Although plasma activation supplies reactive carbon species, thermal energy remains necessary to overcome kinetic barriers associated with carbon diffusion, edge reconstruction, and defect healing. The reduced GNWs coverage observed at 450 °C therefore indicates a significant limited surface diffusion and, consequently, non-uniform coverage despite the plasma assistance that is insufficient to fully compensate for the lower substrate temperature, particularly when conformal coating of complex three-dimensional nanostructures is required.

Transmission Electron Microscopy provides additional information about structural and compositional evidence of the capability of the PECVD process to achieve a conformal coating of GNWs on three-dimensional nanostructures composed of different materials (**Figure 2).** Panel a) shows a high-resolution TEM (HRTEM) image of pristine GNWs, revealing the characteristic vertically oriented, few-layer graphene sheets with wrinkled and corrugated morphology. The thin, semi-transparent contrast and visible lattice fringes confirm the graphitic nature of the nanowalls and serve as a reference for comparison with the hybrid architectures. Panels b) and c) present TEM micrographs of GNWs grown on $SiO_2$ and $TiO_2$ nanotubes ($SiO_2$ NTs@GNWs and $TiO_2$ NTs@GNWs, respectively). In both cases, GNWs are observed to uniformly decorate the entire tip of the nanotubular scaffolds, following their curvature and extending radially outward (in agreement with the previous results in Figure 1 c) and f)). The absence of preferential growth directions or uncoated regions for the $SiO_2$ NTs (Figure 2b) demonstrates that the plasma-assisted growth over vertically aligned samples enables homogeneous coverage even on high-aspect-ratio and curved surfaces. For more intricated scaffolds, as for example, $TiO_2$ NTs self-shadowing effect is observed and nanowall growth is preferentially observed at the top of the nanotube but growing in the same way (Figure 2c). Notably, TEM confirm SEM images where the GNWs preserve their vertical orientation relative to the local surface normal of the nanotubes, indicating surface-energy-driven growth rather than substrate crystallography control (see also Figure S4 for top-view details). Figure 2d) shows the compositional line profile obtained across a single graphene-coated $SiO_2$ nanotube. The C signal is concentrated at the outer region of the nanotube, while Si and O dominate the core, confirming that GNWs form a continuous external shell surrounding the $SiO_2$ nanotube. The gradual transition between signals indicates intimate interfacial contact rather than discrete or delaminated layers. Panels 3 e-f) in Figure 2 display HAADF-STEM images of $SiO_2$NTs@GNWs and $TiO_2$NTs@GNWs, respectively, combined with elemental mapping. The high-angle annular dark-field contrast highlights the dense oxide cores, while carbon mapping clearly outlines a continuous GNWs coating enveloping the nanotube surface. For $TiO_2$NTs, the GNWs layer is seen to extend uniformly over the entire nanotube diameter, further confirming the conformal nature of the coating across different oxide chemistries. Panel g) presents the compositional line profile of $TiO_2$NTs@GNWs, showing a complementary distribution of Ti, O, and C signals. As in the $SiO_2$ case, carbon is predominantly localized at the nanotube exterior, while titanium and oxygen signals are confined to the core. **Figure S8** shows lineal profile for $Al_2O_3$ NTs@GNWs with alike results. Panels h-i) reveal a dense, interconnected GNWs network conformally wrapping the $Al_2O_3$ nanotubes. The high-resolution image (Figure 2 j) shows again a few-layer graphene with short-range ordering and characteristic interlayer spacing, further confirming the graphitic nature of the coating. Importantly, despite the insulating and chemically distinct nature of $Al_2O_3$ compared to $SiO_2$ and $TiO_2$, the GNWs morphology and coverage remain consistent, underscoring the versatility of the PECVD approach. This consistent spatial segregation across different substrates confirms that GNWs growth is independent and governed by plasma-driven surface processes.

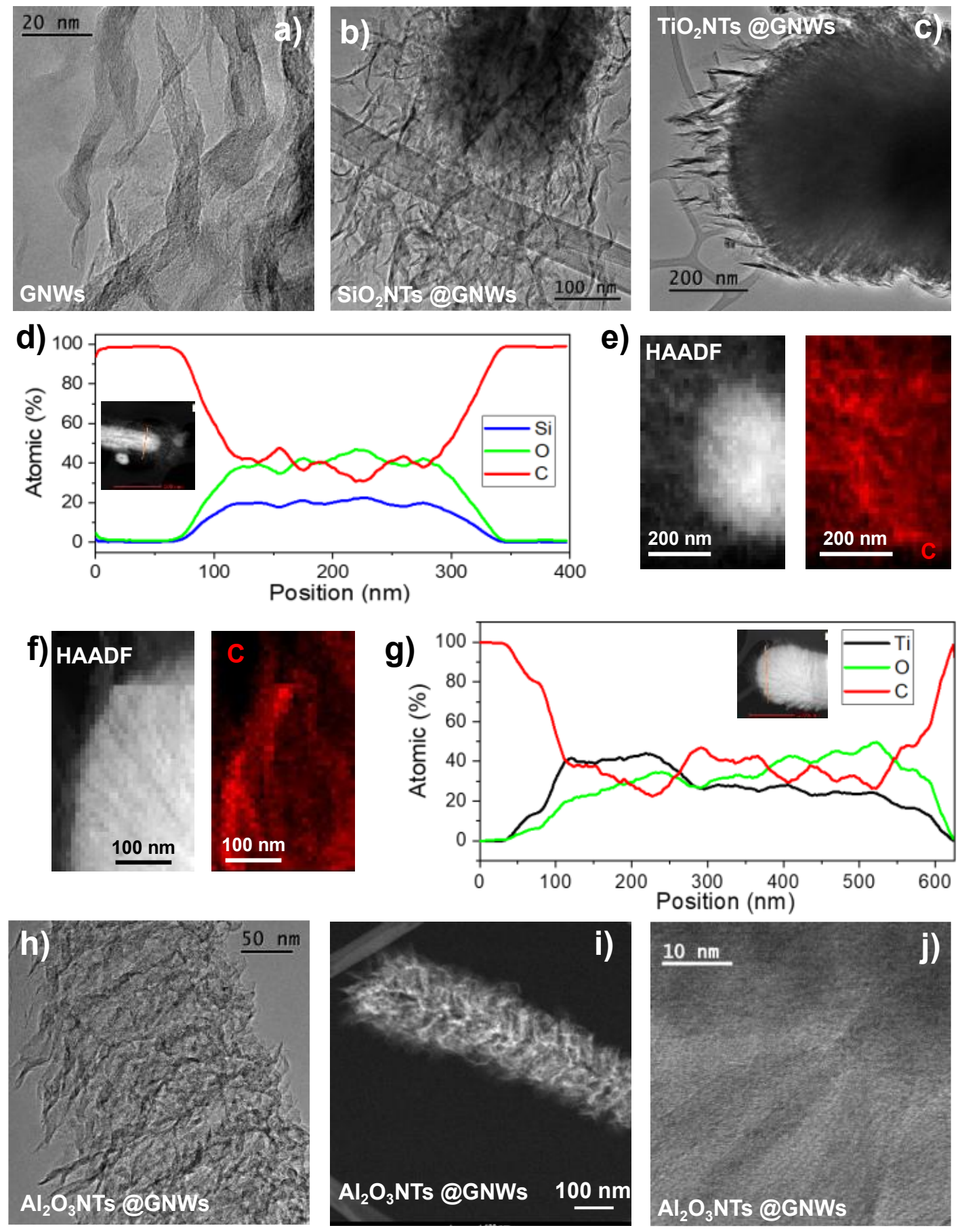

**Figure 2.** a) HRTEM of graphene nanowalls (GNWs), b-c) TEM micrographs of $SiO_2$NTs@GNWs and $TiO_2$NTs@GNWs, d) compositional profile of GNWs@$SiO_2$NTs, e-f) HAADF-STEM micrographs of $SiO_2$NTs@GNWs and $TiO_2$NTs@GNWs, g) compositional profile of $TiO_2$NTs@GNWs, h-i) TEM images of single $Al_2O_3$NTs@GNWs and j) zoomed HRTEM image of $Al_2O_3$NTs@GNWs.

To confirm this finding, we have also performed Raman spectroscopy on all the samples (**Figure 3**). In particular, Figure 3a) shows the normalized Raman spectra of GNWs grown on Si, $SiO_2$, $TiO_2$, and $Al_2O_3$ nanostructured substrates. Raman spectroscopy is a widely established technique for characterizing edge-rich, vertically oriented graphene architectures, as their predominantly $sp^2$-bonded carbon networks produce vibrational fingerprints highly sensitive to lattice order, edge configuration, and defect presence. [60], [61],[62] On a flat substrate, over silicon waffer (red line in panel a), the spectra revealed the fingerprint of typical vertically oriented graphene nanowalls displaying the characteristic D, G, D′, and 2D features of edge-rich, turbostratic graphene structures. In all samples, the D band (at ~1350 $cm^{-1}$) is the most intense peak, indicating the high density of edge-activated scattering pathways associated with vertical graphene nanosheets. This observation is consistent with the mentioned literature, [63], [64], [65] which reports that graphene synthesized by chemical vapor deposition (CVD) and plasma-assisted processes is predominantly characterized by chair edges.

A key factor governing the resulting morphology is the plasma-driven growth mechanism, in which ion bombardment and local electric fields within the plasma sheath preferentially etch $sp^2$ carbon parallel to the substrate while stabilizing edge-terminated graphene sheets, thereby promoting their vertical alignment. Simultaneously, the continuous generation of dangling bonds at GNWs edges sustains further carbon incorporation, enabling self-propagating vertical growth, what accounts for the radial orientation of GNWs across the $MeO_x$ nanotube scaffolds (Figures 1, 2). Despite the rods being vertically aligned, graphene always nucleates and self-organises perpendicular to the local oxide surface: vertically upwards on the rod tips and radially on the rod sidewalls as shown in Figures 2, S4, S5 and S8. Consequently, nanosheets located on the sidewalls appear horizontally aligned relative to the incident laser beam, producing variations in the D′ shoulder (~1610 $cm^{-1}$) and in peak intensity ratios. These differences arise from changes in edge exposure and optical probing geometry [66], [67], rather than from intrinsic crystallinity differences. The 2D band (~2700 $cm^{-1}$) is present in all the samples, indicating few-layer turbostratic stacking with weak interlayer coupling [63], [64], [65]. On the flat Si substrate, the 2D band appears taller and relatively narrower due to direct probing of well-aligned nanosheets in focus, whereas on the oxide nanorod scaffolds, the complex 3D topography and perpendicular nanosheet orientation restrict efficient π-band sampling, causing a broadened and attenuated 2D response (see Figure 3 b) for the shift of D' position and D FWHM through different nanostructures).

Figure 3 c) shows the variation in $I_D/I_G$ and $I_{2D}/I_G$ ratios for each substrate type. GNWs/Si sample displays a relatively tall and narrow 2D band and the highest $I_{2D}/I_G$ value due to the planar geometry that provides efficient optical access to graphitic basal planes. In contrast, when GNWs conformally coat vertically aligned oxide nanotubes, the complex 3D topology reduces direct π-band sampling and slightly increases orientation-induced strain, which broadens and attenuates the 2D response while slightly increasing the D-band linewidth. Although $I_D/I_G$ ratio is commonly used to estimate the defect density in planar graphene, this metric cannot be directly applied to vertically aligned systems. In vertically GNWs, more edges are exposed to the incident Raman laser spot, which in turn gives rise to a sharp D peak. The observed decrease in $I_D/I_G$ values for GNWs on nanostructured substrates can be attributed to the enhanced alignment of the nanowalls with the incident laser, a direct consequence of the nanostructured topography of the substrate [53], [68]. Notably, the high $I_D/I_G$ values observed across all samples primarily reflect edge density and polyaromatic carbon concentration, rather than bulk defects; thus, the conventional relationship for determining the crystallite size (La) cannot be applied to vertically aligned graphene [69], [70]. The $I_{2D}/I_G$ ratio is used to determine the film thickness and number of layers in vertically grown graphene. A high $I_{2D}/I_G$ ratio indicates high-quality structure in flat graphene, but different GNWs samples with a multilayer nature can not be observed in the same way. Overall, the high $I_D/I_G$ ratios, the visible D′ contribution and the broadened 2D response are consistent with the presence of conformal, edge-dominated, vertical graphene networks. The angular distribution and orientation dependence of these networks enhances interfacial charge transfer and opens porosity across the hybrid nano-architecture.[53], [67], [71]

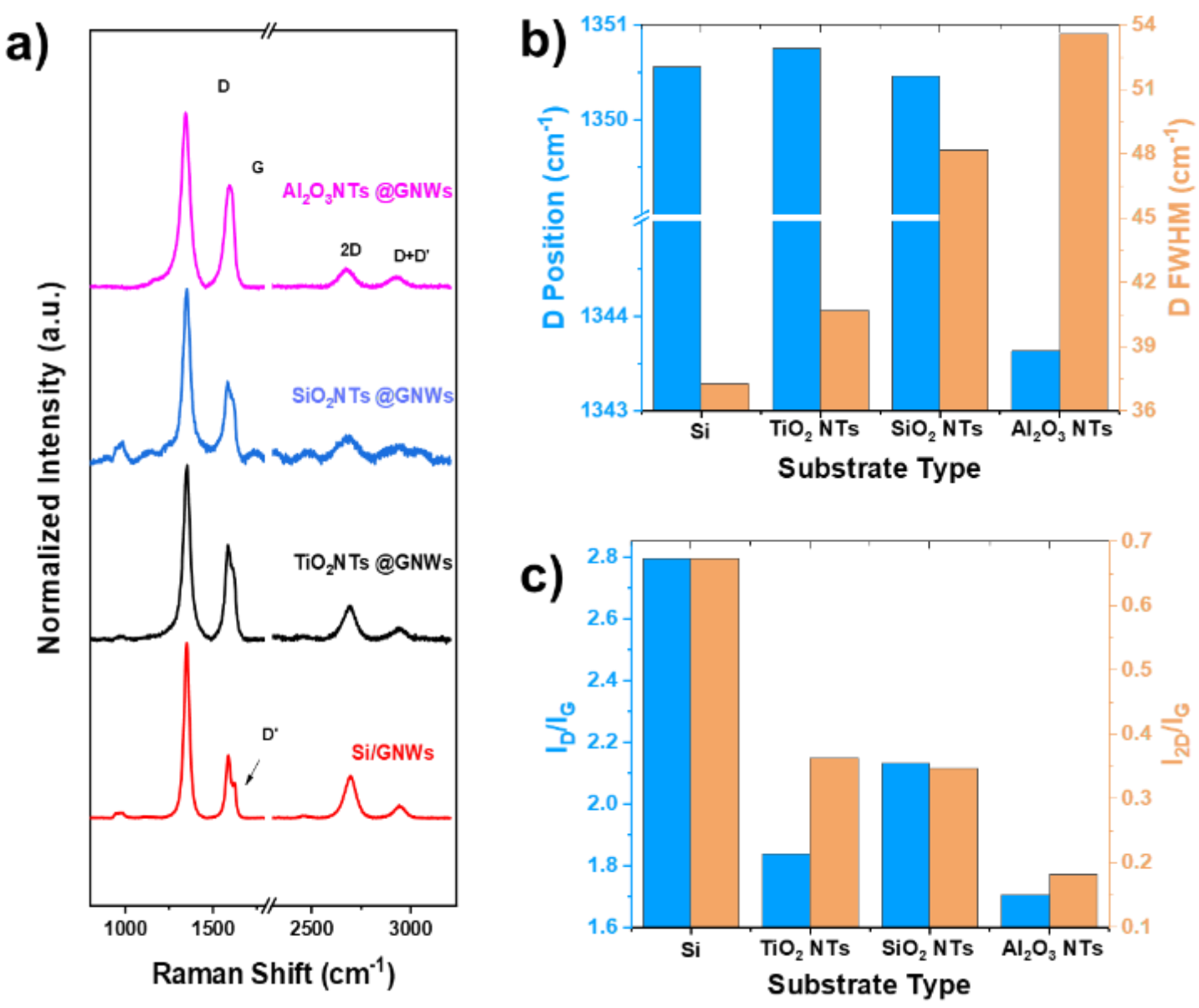

**Figure 3.** a) Raman spectra of all graphene coated nanostructured supports: $TiO_2$ NTs@GNWs, $SiO_2$ NTs@GNWs and $Al_2O_3$ NTs@GNWs, and on the Si reference substrate for comparison. b) D band position in $cm^{-1}$ and D shape FWHM for all VG over different substrates. c) Corresponding $I_{2D}/I_G$ and $I_D/I_G$ ratios for all the explored configurations.

Many factors affect the quality of graphene nanowalls. **Figure S9** presents a full parameterization of GNWs deposition parameters in order to determine their effect on the development of the nanowalls. Figure S9 a) shows how the $I_D/I_G$ and $I_{2D}/I_G$ ratios increase with the plasma power, reaching the best conditions at 30W, corresponding to 2.38 and 0.67 ratios, respectively. The growth temperature also strongly influences graphene quality (Figure S9 b). At 400 °C, the D and G peaks appear almost overlapped, and their intensity ratio indicates the absence of graphene nanosheet formation, instead revealing a carbon film with low structural order. Above 500 °C, and particularly at 600 °C, the G band shifts toward lower wavenumbers, allowing a clear distinction of the S' band. This behavior indicates a higher degree of structural order and is consistent with the literature on $sp^2$ carbon-based materials. Additional factors such as the electrode spacing and the laser incidence angle also contribute to these spectral variations.

**Figure 4** shows the XPS analysis performed to evaluate the chemical composition and surface quality of GNWs grown over different nanostructures ($SiO_2$ NTs, $TiO_2$ NTs and $Al_2O_3$NTs), benchmarked against a flat Si/GNWs reference. Survey spectra for the three architectures reveal in **Figure S10** a marked carbon increase after GNWs deposition, consistent with conformal coverage of the nanostructured supports. A basic quantification of the Ti/C, Si/C and Al/C ratios before and after the GNWs growth (**Table S1** in Supporting Information) confirms the virtually continuous carbon coverage on all the nanostructured supports. Residual signals of titanium and aluminium suggest either higher GNWs porosity or less uniform GNWs coverage compared to the case of the $SiO_2$ NTs substrate. Nevertheless, graphene grown on $SiO_2$ NTs presented the higher non-oxidized oxygen / carbon ratio compared to graphene developed on $TiO_2$ and $Al_2O_3$ NTs. Although being the half of the O/C values reported in the literature for the development of graphene oxide (~0.4-1.2)[72], the oxygen content on the $SiO_2$ NTs@GNWs sample indicates a significantly oxygen-functionalized graphene surface. In contrast the much lower ratios observed for $Al_2O_3$ and $TiO_2$ NTs substrates indicate a predominantly graphitic character with limited residual oxygen functionalities at the surface. The observed trend is consistent with the interfacial chemistry of the underlying oxides: $SiO_2$ surfaces are rich in silanol (Si–OH) groups that promote hydrogen bonding and stabilization of oxygen-containing groups on graphene, promoting the retention of GO-like functionalities. Conversely, $Al_2O_3$ and $TiO_2$ surfaces tend to form stronger covalent interfacial bonds (e.g., C–O–M linkages)[73] , which can reduce the amount of physisorbed or weakly bound oxygen species. Deconvolution of the C1s spectra (Figure 4a) further clarifies these differences. For the Si/GNWs reference only three components are found, C=C ($sp^2$) at 284.5 eV, C-C/C-H ($sp^3$) at 285.0 eV and C-OH/ C-O-C bounds at 286.3 eV. [74], [75]. The lower intensity of higher binding energy components confirms the high graphitic quality of the GNWs grown by PECVD, with oxygen functionalities (~3.7%) attributable mainly to adsorbed moisture and adventitious carbon contaminants. In contrast, GNWs grown on nanostructured oxides supports (panels 4 b-d for GNWs grown on $TiO_2$, $SiO_2$, and $Al_2O_3$ nanotubes, respectively) exhibit a clear photopeak tail at high binding energies. This can be linked to the presence of additional contributions such as C=O, O-C=O and π-π* bonds at 288.0 eV, 289.1 eV and 291.0 eV respectively. [74], [75] Diagram bar in **Figure S11** summarizes the relative contribution of the different carbon functionalities. Although $sp^2$ carbon remains dominant in all samples, its fraction decreases systematically on nanostructured supports, consistent with curvature-induced strain, increased edge density and growth-induced defects that promote the formation of $sp^3$ bonds and reactive anchoring sites. [60], [73] The observed trend in $sp^3$ content among the supports ($SiO_2$< $Al_2O_3$ < $TiO_2$) correlates with a the increasing polarity and and density of unsaturated surface states of the oxides, which enhance graphene-substrate interactions and facilitate chemical functionalization. [76] Additionally, the small presence of oxidized groups (C–O, C=O, O–C=O) and π–π* transitions in GNWs on nanostructured supports reflects the higher density of edges and exposed defect sites, which acts as adsorption centers for oxygenated species during or after growth. The extracted $sp^2/sp^3$ ratio obtained with the C=C/C-C contributions of the C 1s fitting had values like 4.5, 3.3 and 2.6 for $SiO_2$ NTs@GNWs, $Al_2O_3$ NTs@GNWs and $TiO_2$ NTs@GNWs respectively, compared to the 3.75 value corresponding to the GNWs on a flat reference, aligned with literature reports showing that metal oxides can induce structural disorder and interfacial roughness, increasing apparent $sp^3$-like contributions even when the basal graphene lattice remains largely continuous. [62] Moreover, specific oxide seed can promote distinct defect types: Al-oxides may favor C–O and C–Al related $sp^3$ defects, which reduce the apparent $sp^2$ fraction relative to Ti-oxides. [76] Meanwhile, hydrophilic $SiO_2$ surfaces can introduce localized deformation and additional $sp^3$-like without fundamentally altering the extended $sp^2$ network. [55] This XPS trends are fully compatible with the Raman $I_D/I_G$ values, as samples with similar $sp^2/sp^3$ ratios may exhibit different Raman defect intensities depending on defect type, spatial distribution, substrate interaction, and local strain fields. [62]

To complement the chemical analysis, Near Edge X-ray Absorption Fine Structure Spectroscopy (NEXAFS) is used to probe the chemical bonding, structural and orientation information of GNWs. Figure 4e) presents C K-edge spectra of GNWs deposited on flat Si substrate (100) and on different nanostructured supports. NEXAFS was performed to probe the chemical composition and structural orientation of the grown graphene on the surface. The C K-edge spectra were obtained in partial electron yield (PEY) mode with a retarding potential of -150 V, enabling the top surface analysis. Spectra presented herein are obtained at 55°, i.e., magic angle. Raw spectra were treated as described elsewhere. [53], [54] The spectra of all samples exhibit similar features and have been normalized to the double peak at 292.12 eV, for easier comparison. All recorded spectra, whether from flat or nanotube-based surfaces, measured at the top surface display nearly identical features, characteristic of vertically grown graphene. As described previously, they are defined by two distinct signatures: a resonance at ~285 eV corresponding to the C 1s → π* transition, and a doublet of core-excitonic origin around 292 eV, followed by broader band-like contributions spanning 293–330 eV associated with C 1s → σ* transitions.

This σ* double resonance around 292 eV is composed of a sharp and a broad feature. The first sharp feature is generally identified as a core excitonic. Its sharpness reflects the strong correlation of electron-hole pairs within the material. The second feature of σ* resonance is related to the transition from the 1s level to relatively non-dispersing σ* states near the Γ point of the BZ. It is a more delocalized contribution. Such configuration in double exciton features is typical for graphene-like structures and reveal intact several layer graphene with a few defects. Interesting detail is that we do not observe strong presence of foreign atoms during the analysis of carbon layers at C K-edge. This would attribute the oxygen in XPS to inner part of the 2D/3D structures.

To investigate in detail the orientation of the electronic orbitals, angle-resolved NEXAFS measurements were performed at incidence angles of 30°, 55°, and 90° of a linearly polarized synchrotron X-ray beam on 2D, an on GNWs on 3D surfaces (**Figure S12**). Analysis revealed that the π* resonance showed angular dependency. Its intensity increased at higher incidence angles and decreased at lower angles (30°). In contrast, the σ* band shows the reverse trend, validating the vertical orientation of the grown graphene walls. In GNWs, the graphene sheets are standing perpendicular to the substrate surface. The intensity of NEXAFS resonance is maximized when the electric field E-vector of the incident polarized X-ray beam is aligned parallel to the direction of the final state orbital. Orientation changes the angular dependence of spectra in horizontal graphene materials. In vertical graphene, the π* lies mostly parallel to the substrate surface plane. Consequently, the π* resonance signal's intensity is maximized when the incident E-vector is also parallel to the substrate surface. Conversely, the σ* orbitals lie in-plane to the vertical graphene sheets, meaning they are mostly perpendicular to the substrate surface plane. Thus, when the incident E-vector is parallel to the substrate surface, the intensity of σ* resonance is minimized. Angle resolved data showed difference between 2D and 3D substrates: graphene clearly vertically oriented on 2D, whereas the orientation changes slightly toward horizontal in the 3D samples, describing the presence of the walls that are normal to the vertical sides of the tubes.

It is very clear that the character of the nanowalls stays same in all of the directions (angle dependent NEXAFS changes in sp2 were the final proof for this assumption.

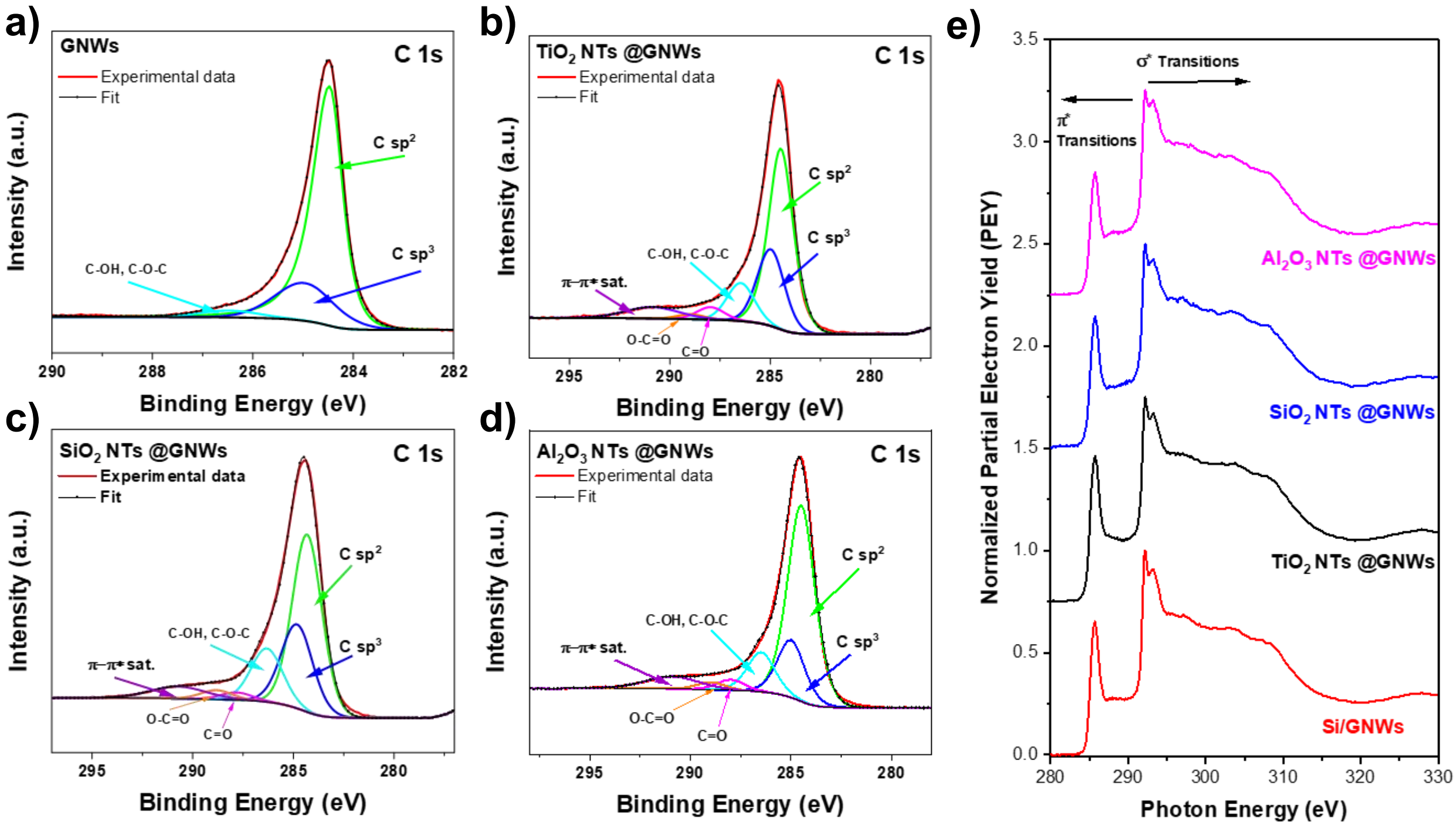

**Figure 4.** XPS chemical composition analysis of the carbon functionalities through the C1s spectra at the surface of a) Si/GNWs, b) $TiO_2$ NTs@GNWs, c) $SiO_2$ NTs @GNWs and d) $Al_2O_3$ NTs @GNWs, e) NEXAFS spectra for GNWs over the samples already mentioned.

### 3.3 Superwettability on $MeO_x$ @GNWs structures.

Wettability is a key parameter for the targeted applications of these nanostructured surfaces. **Table S2** summarizes the main wetting parameters obtained from standard wetting measurements and controlled-atmosphere condensation tests. The wetting behavior of GNWs grown on a flat reference is fully consistent with the developed nanoarchitecture. Although the surface exhibits an apparent WCA >130º with strong pinning (CAH ~20º and no sliding of 10µl water droplets), this response cannot be attributed to a graphene hydrophobicity supported by its chemical nature but rather to morphology. A rough estimation of the internal Laplace pressure of a deposited macroscopic water droplet onto the GNWs surface is extremely small (~ 100 Pa) compared to the tens to hundreds of kPa require to force penetration into GNWs nanocavities of 50-200 nm, as established in classical capillary-pressure stability analysis.[77], [78] This mismatch implies that the droplet does not have sufficient driving force to fully impregnate the nanopores and therefore the nanostructured surface remains in a metastable Cassie-Baxter configuration. [79] The measured WCA can instead be rationalized by assuming and intrinsic WCA ~80º and a realistic effective solid fraction (φ~0.22),[80], [81] supporting a sticky (strong adhesion) hydrophobic behavior dictated by the graphene nanostructuration (including possible defects or oxygenated functionalities), where suspended regions coexist with partial penetration in vertical or re-entrant nanocavities.

Surfaces consisting of GNWs deposited on either superhydrophilic ($SiO_2$) or superhydrophobic ($TiO_2$, $Al_2O_3$) supports exhibit superhydrophobic behavior, with WCAs > 175º, water contact angle hystheresis < 10º, and rolling-off angles < 5º. The theoretical solid fraction required to reproduce the measured WCA is very small (φ~0.02) being fully consistent with a hierarchical Cassie-Baxter wetting state. **Figure 5** a) illustrates the low-adhesion character of $TiO_2$ NTs@GNWs, where water droplets bounce and slide off the surface, leaving it clean and dry.

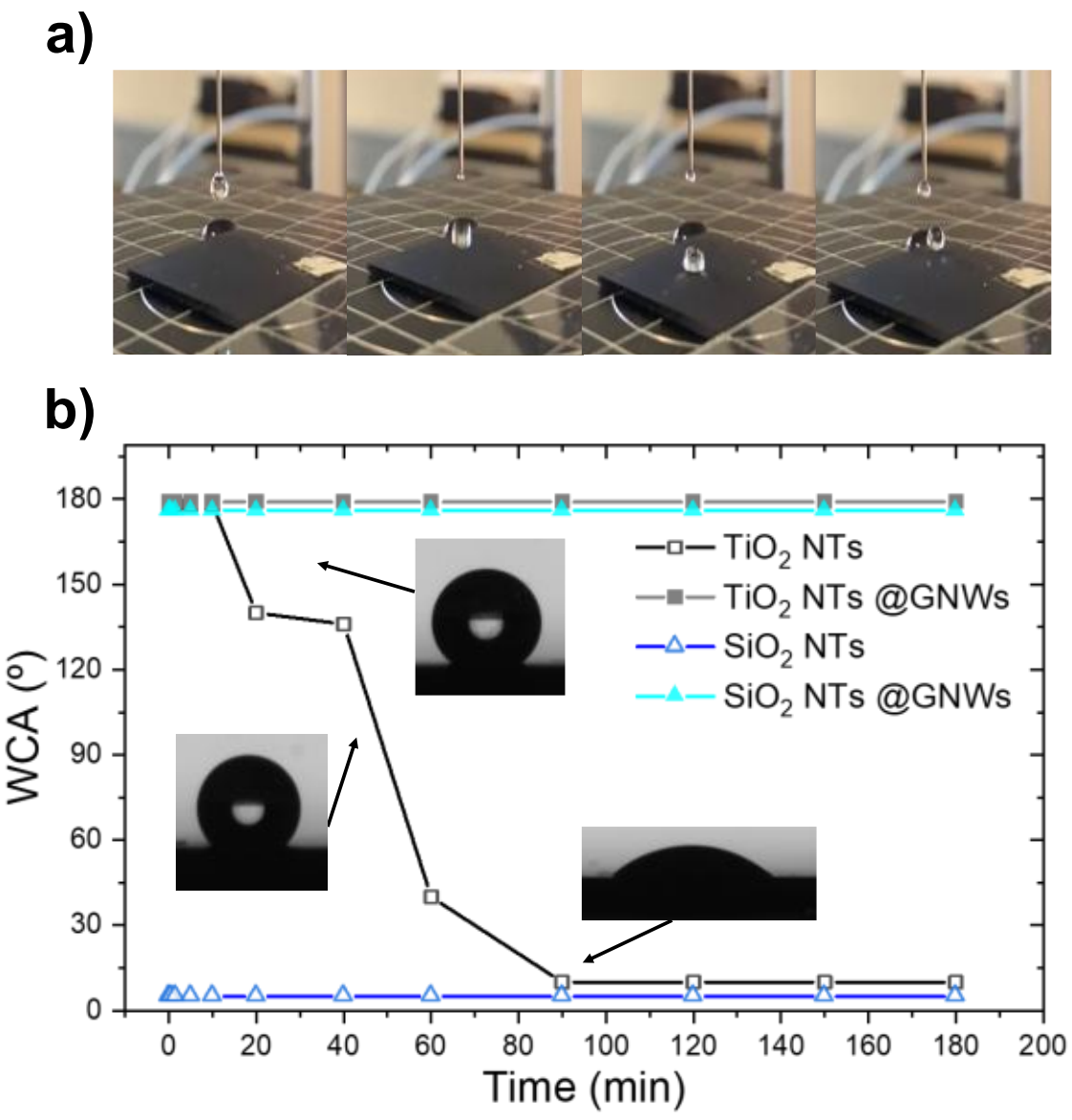

**Figure 5.** a) Sequence of images of the $TiO_2$NTs@GNWs sample demonstrating a superhydrophobic behavior through the bouncing of 5 µL water droplet that rolls off the surface. b) WCA of 2 µL evolution of different hierarchical surfaces after UV-vis light irradiation under room conditions for the indicated times with a 200W UV lamp at 30 cm of distance incident directly over 90° tilted samples.

The 3D $MeO_x$ NTs@GNWs architecture provides a multiscale roughness ranging from sub-nanometer (graphene nanowalls) to micrometer (nanotube length). Such hierarchical and partially re-entrant morphology is known to promote superomniphobicity by stabilizing air pockets and suppressing liquid impalement. The fact that all GNWs-decorated nanostructured surfaces evolve toward superhydrophobicity highlights the versatility of the vacuum-based fabrication route, which is independent of the underlying metal oxide chemistry and compatible with a wide variety of 3D substrates.

To evaluate the stability of the superomniphobic state, we performed a UV-vis photoactivation assay on all graphene-based hierarchical surfaces and compared them with the bare nanostructured supports, irradiating with a Xe lamp for increasing exposure times (Figure 5 b). As expected from the intrinsic photoactive nature of $TiO_2$ NTs, the uncoated scaffolds undergo a reversible transition from an initial superhydrophobic state (WCA >175º) to a superhydrophilic one (WCA<10º) upon UV illumination. This behavior corresponds to the well-known transition from a metastable Cassie-Baxter regime to a fully wetted Wenzel one [30] triggered by UV-induced surface hydroxylation. In the pristine state, air trapped within the 3D nanostructure pores produce dual-scale repellency (assuming densities over 1NT/$\mu m^2$), whereas UV exposure progressively leads to complete wetting of the surface roughness. In contrast, $SiO_2$ NTs remain superydrophilic from the very beginning, as expected for a non-photoactive oxide. The behavior changes markedly when the NTs scaffolds are decorated with graphene nanowalls. UV irradiation does not significantly modify the surface chemistry of the GNW-coated architectures, which maintain high WCAs and low hysteresis throughout the entire exposure period (up to 180 min). The inset images in Figure 5 b) clearly illustrate this transition: while water droplets on bare $TiO_2$ NTs gradually flatten and pin, droplets on hierarchical GNW-coated surfaces remain nearly spherical with minimal adhesion. It is worth noting that the high temperature GNW deposition process converts $TiO_2$ NTs from amorphous to anatase crystalline structure (see Figure S6), which would normally enhance photocatalytic activity and accelerate UV-induced wettability changes. [82] However, the GNWs act as an effective protective agent, suppressing UV-driven hydroxylation of the exposed $TiO_2$ surface and preserving the low-energy interface required for air entrapment. This efficient repellent behavior for sessile drops is consistent with a stable Cassie-Baxter regime supported by the hierarchical topography.

It is worth noting that under realistic operating environments, superhydrophobic surfaces are frequently exposed to condensation phenomena, which can critically compromise their wetting performance. The formation and growth of liquid droplets under high humidity or sub-zero conditions may promote liquid infiltration into the surface texture, triggering a transition from the metastable Cassie–Baxter state to the fully wetted Wenzel state. This transition is particularly detrimental for applications that rely on sustained non-wetting behavior, such as anti-fouling coatings, water harvesting systems, and moisture-driven energy harvesting devices. Therefore, understanding the dynamic interaction between condensed droplets and hierarchical surface architectures under controlled environmental conditions is essential to assess the robustness and functional reliability of these materials.

In this context, environmental SEM (ESEM) provides a powerful approach to directly visualize condensation processes at the microscale and elucidate the mechanisms governing wetting state stability (**Figure 6**). On the left-side (a-d) top-view micrographs present the evolution of $TiO_2$ NTs @GNWs sample stablishing the operation conditions in -5ºC for Peltier's temperature, saturated relative humidity of 100% and modifying the water vapor pressure from 445 to 475 Pa inside the analysis chamber. Initially, isolated, randomly distributed tiny condensed water droplets with smalls diameters, below 10 µm, are detected (Figure 6 b). They increase their average volume until 1 nl for the first 90 seconds (see also **Figure S13**) at the same time as other microdroplets are appearing (Figure 6 c) till they start to join and coalesce forming bigger ones (Figure 5 d), of an average volume around 10 nL at 140 s of the condensation experiment (Figure S13).

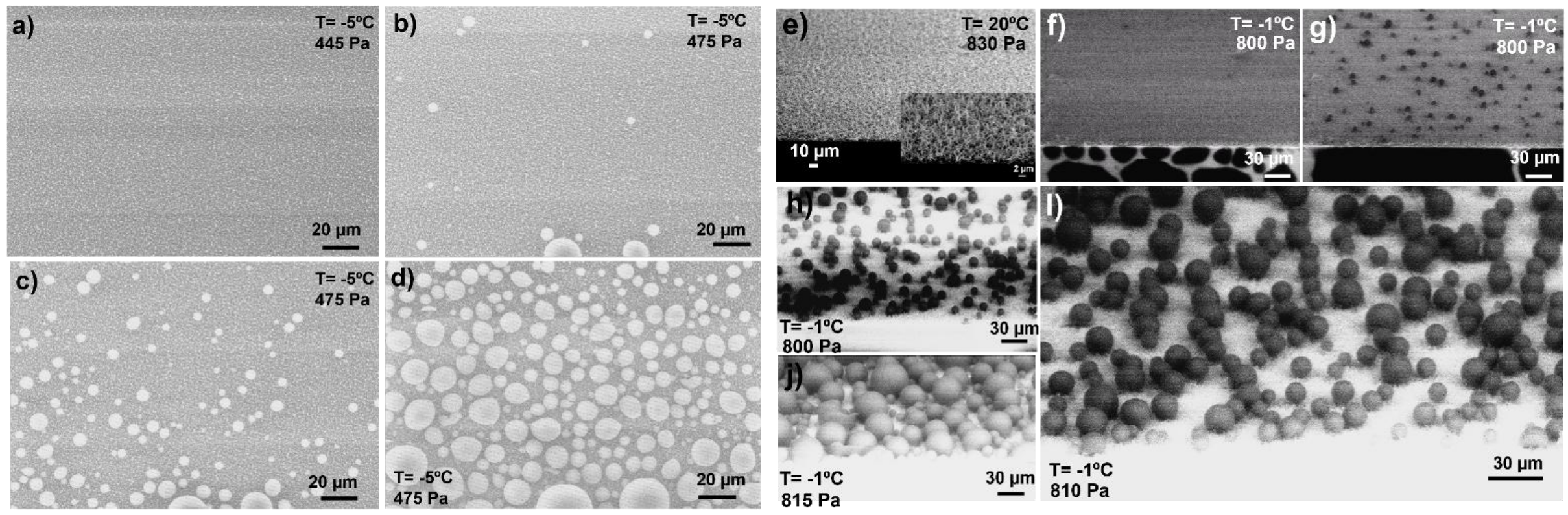

**Figure 6.** a-d) Top-view ESEM micrographs showing the water condensation on the superhydrophobic $TiO_2$ NTs @GNWs surface at -5ºC and 100% RH increasing the water vapor pressure until 475 Pa at time intervals of 20-30 seconds; h-j) Tilted cross-section ESEM micrographs evolution for the same sample at -1ºC and 100% RH increasing the water vapor pressure until 815 Pa at time intervals of 20-30 s (see figure S13 for droplet growth rate observation) .

The drop-shape remains isolated, practically spherical according to a robust superhydrophobic behaviour even with indications of a possible graphene induced pinning effects given the static state of the condensed microdroplets (see Figure S13b). ESEM micrographs of tilted sections acquired at −1 °C (Figures 6 h–j) confirm the previous observation: as condensation progressed onto silicon substrate section (Figure 6 f, g), the hierarchical surface delayed the nucleation phenomenon. Besides it is possible to distinguish how condensed water droplets remained supported by the nanostructures without being absorbed into the condensed water droplet volume (Figure 6 i), indicating a strong resistance to capillary-driven impregnation. These results indicate that the hierarchical combination of materials effectively delays the Cassie–Baxter to Wenzel transition, even under thermodynamically unfavorable conditions.

Finally, **Figure 7** demonstrates that the hierarchical carbon-based nanostructures also repel non-polar and complex organic liquids, reaching apparent contact angles >175º for most simulated fouling fluids. While hierarchical roughness has been widely exploited for anti-fouling, anti-bacterial, self-cleanable, water filtration, anti-icing and water management surfaces, [83] achieving omniphobicity typically requires fluorinated or silane-based functionalization. [84] In contrast, our hierarchical GNW-coated nanostructures achieve marked repellence behavior through a fluorine-free strategy.

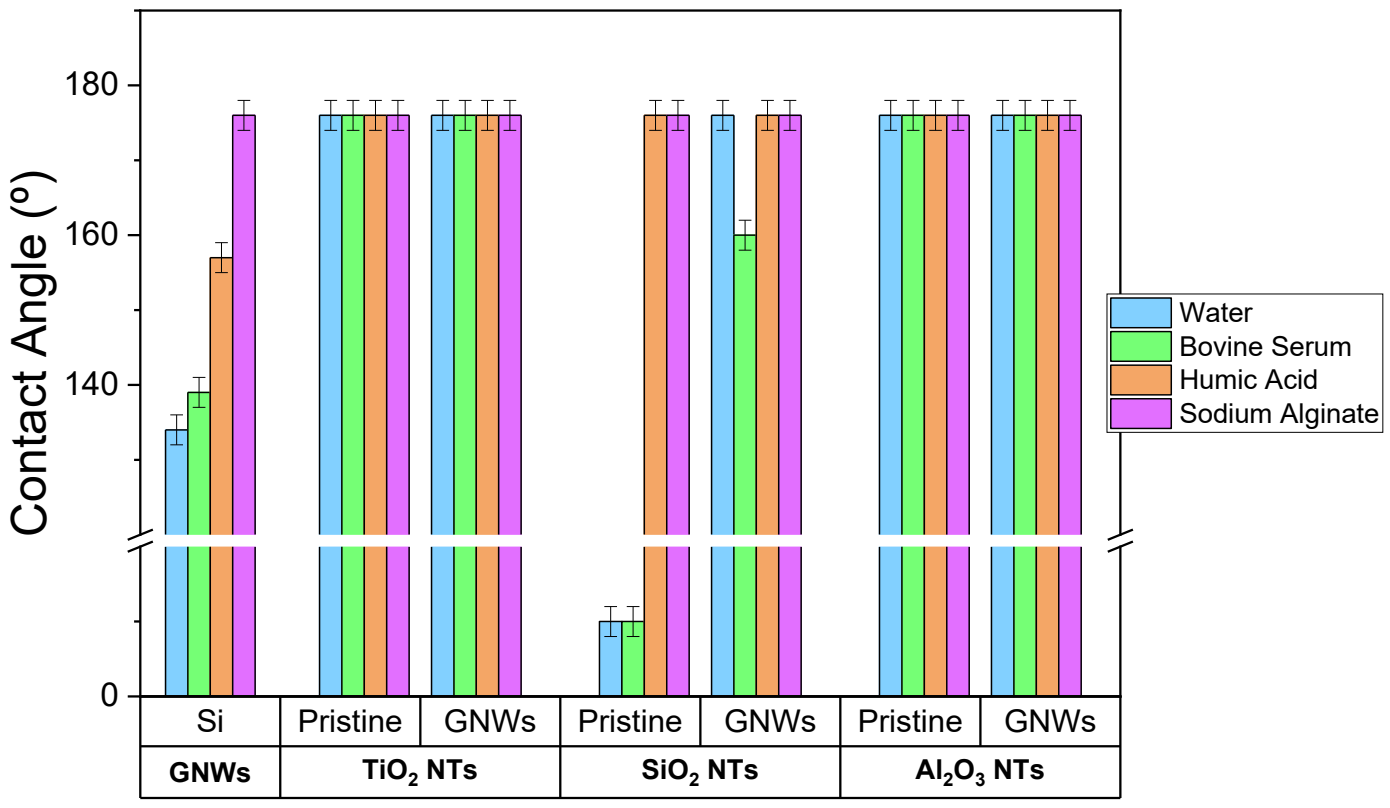

**Figure 7.** Contact angle measurements for different 2 μL fluid droplets (water, bovine serum, humic acid and sodium alginate) for as-grown $MeO_x$ and graphene coated NTs.

Overall, these results demonstrate that the synergistic combination of graphene nanosheets and $MeO_x$ nanotube hierarchical structuring enables a chemically robust, photo-stable, and condensation-resistant omniphobic surface. The sustained Cassie–Baxter wetting state across diverse liquids and environmental conditions highlights the potential of $MeO_x$ NTs @GNWs as sustainable, fluorine-free solution for applications in anti-fouling and biofluid-repellent coatings.

## 4 Conclusions

The combination of soft-templated oxide nanotubes with plasma-assisted fabrication of graphene nanowalls yields a hierarchical, re-entrant 3D architecture that unites high surface aspect ratio scaffolds and structural rigidity with conformal, electrically conductive graphitic networks. Comprehensive structural and spectroscopic analysis confirms the formation of uniformly decorated nanotube arrays, where radially GNWs preserve their $sp^2$ framework independently from substrate nature, and they preserve controlled degree of edge functionalization. This balance between graphitic integrity and interfacial chemical activity is particularly advantageous for applications requiring simultaneous electrical transport and surface reactivity, including electrocatalysis, energy storage and electrochemical sensing.

The multiscale, re-entrant morphology produced by this hybrid design also enables long-term superomniphobic behavior. Graphene nanowalls effectively lower the surface energy and reinforce air entrapment, resulting in stable repellency against chemically diverse liquids. Comprensive evaluation across $TiO_2$, $SiO_2$, and $Al_2O_3$ nanotube systems demonstrates that graphene nanostructures dominate the wetting response, largely decoupling surface behavior from the intrinsic oxide shell chemistry. The resulting surfaces maintain robust omniphobicity and resist wetting transitions under UV irradiation and water condensation, underscoring the generality of this approach for oxide-based functional interfaces.

Overall, this work establishes a substrate-compatible, low temperature (~450 °C) and scalable route to architecturally tunable graphene-oxide heterostructures. Their combined electrical, chemical and wetting functionalities position these strategy as promising candidates next generation microscale functional devices, including3D structure materials requiring durable, high-performance surface engineering.

These hybrid configurations can be compared to those obtained as VG on any $MeO_x$ flat surface, as they offer clear advantages in areas such as electrochemical sensing[86], [87]. Here, the edges of the graphene facilitate rapid charge transfer, while the oxide component promotes selective interactions or catalytic activity. Similarly, in field emission and vacuum microelectronics, the combination of sharp graphene edges and geometric amplification provided by the nanotube scaffold improves emission stability and efficiency. The architecture also offers interesting prospects for optoelectronics and photocatalysis, where the oxide nanostructures contribute to photoresponses and the graphene channels efficiently extract charges. [73]

## Acknowledgements

We thank the projects PID2025-175485OB-I00, PID2022-143120OB-100, PCI2024-153451 and TED2021-130916B-I00 funded by MCIN/AEI/10.13039/501100011033 and by "ERDF(FEDER) A way of making Europe, Fondos Nextgeneration EU and Plan de Recuperación, Transformación y Resiliencia", by the European Union, the EU H2020 program under grant agreement 3Dscavengers Starting Grant 851929 and Junta de Andalucía through the DGP_PIDI_2024_02239 project. Project ANGSTROM was selected in the Joint Transnational Call 2023 of M-ERA.NET 3, an EU-funded network of about 49 funding organizations (Horizon 2020 grant agreement No 958174). FNG thanks to the "VII Plan Propio de Investigación y Transferencia" of the Universidad de Sevilla. The authors also acknowledge the access to the Determination of Surface Wetting, scientific-technical service of the Institute of Materials Science of Seville, funded by Junta de Andalucia (IE19_142 USE I+D+I Infrastructure and Equipment 2019-PAIDI2020 European Regional Development Fund). The authors also acknowledge valuable support by EU Graphene Flagship FLAG-ERA III JTC 2021 project VEGA (ANR-21-GRF1-0004 or MIZS-PR-11938). The authors also want to thank HZB for the allocation of synchrotron radiation beam time at Bessy II at HESGM beam station, and the Region Centre Val de Loire, France, for the support via project APR IR "OPTIMICROCAP" as well as the fellowship provided by Ecole Doctorale EMSTU Orleans

## Data availability

- The data supporting this article have been included as part of the Supplementary Information, and the article will be deposited at repository as ARXIVE and HAL.

.

## Authors' Contributions:

Fernando Núñez-Gálvez: Conceptualization of the hierarchical graphene–oxide architecture, oxide nanotube fabrication, methodology, investigation, structural and wettability analysis, data curation, visualization, writing—original draft, and writing—review and editing.

Muhammad Hamza: Plasma-assisted graphene growth, in situ Raman spectroscopy, materials characterization, formal analysis, data curation, visualization, and writing—review and editing.

Johannes Berndt: Conceptualization of the plasma-assisted graphene-growth strategy, plasma-process methodology, supervision of plasma experiments and in situ Raman analysis, validation, resources, and writing—review and editing.

Thomas Strunskus: synchrotron analysis, surface chemical analysis, formal analysis, validation, resources, and writing—review and editing.

Franziska Traeger: NEXAFS measurements and analysis, surface and electronic-structure characterization, formal analysis, validation, and writing—review and editing.

Nicoló di Novo: Investigation, morphological and wettability characterization, formal analysis, data curation, visualization, and writing—review and editing.

Juan R. Sánchez-Valencia: Oxide nanotube synthesis and characterization, methodology, investigation, validation, resources, supervision, and writing—review and editing.

Carmen Lopez-Santos: Conceptualization and development of the oxide nanotube scaffolds, surface characterization, methodology, formal analysis, supervision, writing—original draft and writing—review and editing.

Ana Borras: Conceptualization of the oxide-based architecture, methodology, resources, supervision, project administration, funding acquisition, writing—original draft and writing—review and editing.

Eva Kovacevic: Conceptualization of the plasma-assisted conformal graphene-growth approach for 2D 3D architectures, methodology, supervision of graphene growth and characterization, drafting, resources, project administration, funding acquisition, writing—original draft and writing—review and editing.

All authors contributed to the interpretation and discussion of the structural, spectroscopic, surface-chemical and wettability results, reviewed the manuscript, and approved the final version.

## Declarations

**Conflict of Interest** The authors declare no interest conflict. They have no known competing financial interests or personal relationships that could have appeared to influence the work reported in this paper.

### Figure captions

**Scheme 1.** Step-by-step formation of $MeO_x$ @GNWs enabled by vacuum and plasma processes.

**Figure 1. Microstructure of the 3D Hierarchical $MeO_x$@GNWs supported nanostructures.** Characteristic SEM micrographs showing the formation of GNWs on supported $MeO_x$ nanotubes on Si wafers for: a) as-grown $SiO_2$ nanotubes, b-c) $SiO_2$ NTs@GNWs (i.e., after deposition of GNW), d) $TiO_2$ NTs, e-f) $TiO_2$ NTs@GNWs, g) detailed

top-view of a graphene-coated single $TiO_2$ nanotube, h) $Al_2O_3$ NTs before GNWs deposition, i-j) $Al_2O_3$ NTs@GNWs and k) top-view of the previous sample.

**Figure 2.** a) HRTEM of graphene nanowalls (GNWs), b-c) TEM micrographs of $SiO_2$NTs@GNWs and $TiO_2$NTs@GNWs, d) compositional profile of GNWs@$SiO_2$NTs, e-f) HAADF-STEM micrographs of $SiO_2$NTs@GNWs and $TiO_2$NTs@GNWs, g) compositional profile of $TiO_2$NTs@GNWs, h-i) TEM images of single $Al_2O_3$NTs@GNWs and j) zoomed HRTEM image of $Al_2O_3$NTs@GNWs

**Figure 3.** a) Raman spectra of all graphene coated nanostructured supports: $TiO_2$ NTs@GNWs, $SiO_2$ NTs@GNWs and $Al_2O_3$ NTs@GNWs, and on the Si reference substrate for comparison. b) D band position in $cm^{-1}$ and D shape FWHM for all VG over different substrates. c) Corresponding $I_{2D}/I_G$ and $I_D/I_G$ ratios for all the explored configurations.

**Figure 4.** XPS chemical composition analysis of the carbon functionalities through the C1s spectra at the surface of a) Si/GNWs, b) $TiO_2$ NTs@GNWs, c) $SiO_2$ NTs @GNWs and d) $Al_2O_3$ NTs @GNWs, e) NEXAFS spectra for GNWs over the samples already mentioned.

**Figure 5.** a) Sequence of images of the $TiO_2$NTs@GNWs sample demonstrating a superhydrophobic behavior through the bouncing of 5 μL water droplet that rolls off the surface. b) WCA of 2 μL evolution of different hierarchical surfaces after UV-vis light irradiation under room conditions for the indicated times with a 200W UV lamp at 30 cm of distance incident directly over 90° tilted samples.

**Figure 6.** a-d) Top-view ESEM micrographs showing the water condensation on the superhydrophobic $TiO_2$ NTs @GNWs surface at -5ºC and 100% RH increasing the water vapor pressure until 475 Pa at time intervals of 20-30 seconds; h-j) Tilted cross-section ESEM micrographs evolution for the same sample at -1ºC and 100% RH increasing the water vapor

pressure until 815 Pa at time intervals of 20-30 s (see figure S13 for droplet growth rate observation) .

**Figure 7.** Contact angle measurements for different 2 μL fluid droplets (water, bovine serum, humic acid and sodium alginate) for as-grown $MeO_x$ and graphene coated NTs.

# Supporting Information

## Tunable Conformal Graphene Growth on Oxide Nanotube scaffolds: Towards superwettable hierarchical 2D–3D Architectures

Fernando Núñez-Gálvez[1,2,*], Muhammad Hamza[3], Johannes Berndt[3], Thomas Strunskus[4], Franziska Traeger[5], Nicoló Di Novo,[2] Juan R. Sánchez-Valencia[2], Carmen Lopez-Santos[1,2], Ana Borras[2,*] and Eva Kovacevic[3,*]

[1] Departamento de Física Aplicada I, Escuela Politécnica Superior, Universidad de Sevilla

[2] Nanotechnology on Surfaces and Plasma Lab, Institute of Materials Science of Seville (CSIC-US)

[3] GREMI, Groupe de Recherches sur l'Energétique des Milieux Ionisés (CNRS-Université d'Orléans)

[4] Department of Materials Science, Kiel University, Kaiserstr, 2, 24143 Kiel, Germany

[5] Westfälische Hochschule, August-Schmidt-Ring 10, 45665 Recklinghausen, Germany

*fngalvez@us.es; fernando.ngalvez@icmse.csic.es, anaisabel.borras@icmse.csic.es; eva.kovacevic@univ-orleans.fr

**Figure S1.** Cross-sectional SEM micrographs for the seed thin film a) $TiO_2$ and b) $SiO_2$ prepared for the formation of the 3D nanostructures

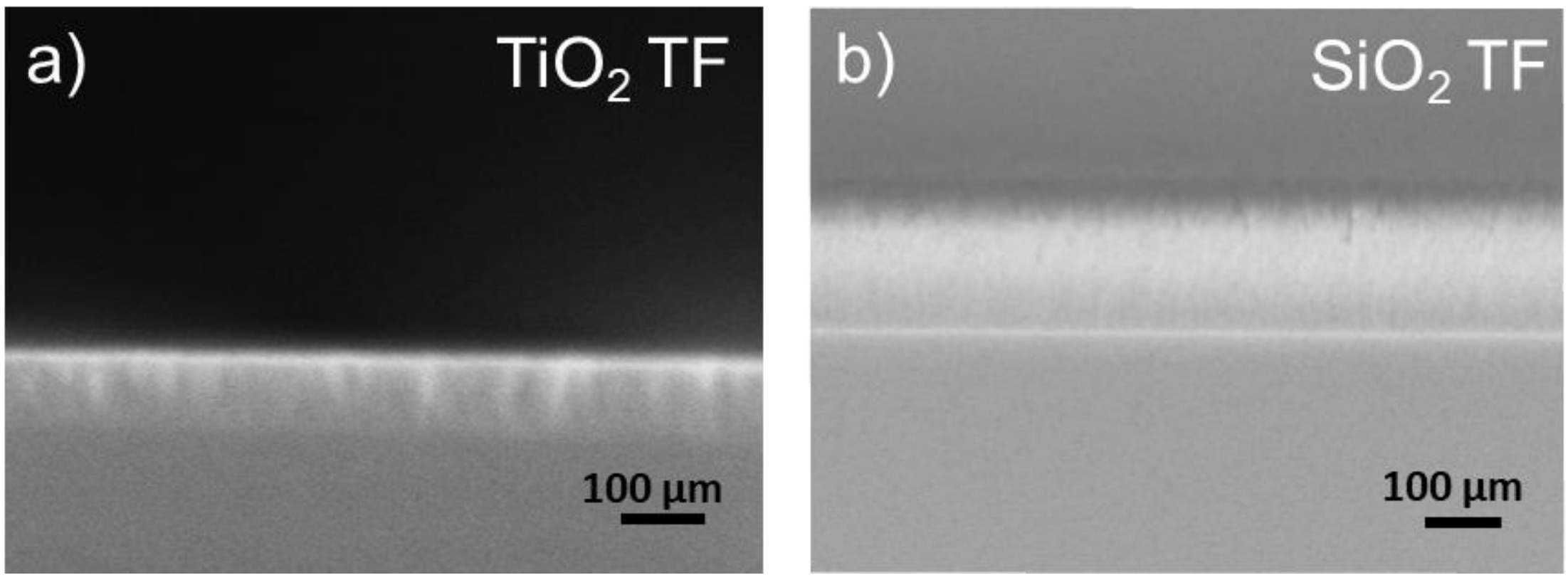

**Figure S2.** Statistical analysis of number and length of the nanotubes for different samples ($SiO_2$ NTs, $TiO_2$ NTs and $Al_2O_3$ NTs). A minimum of 100 single nanotubes (for length and diameter analysis) were characterized using ImageJ software. Nanotube density was also calculated using at least 5 images.

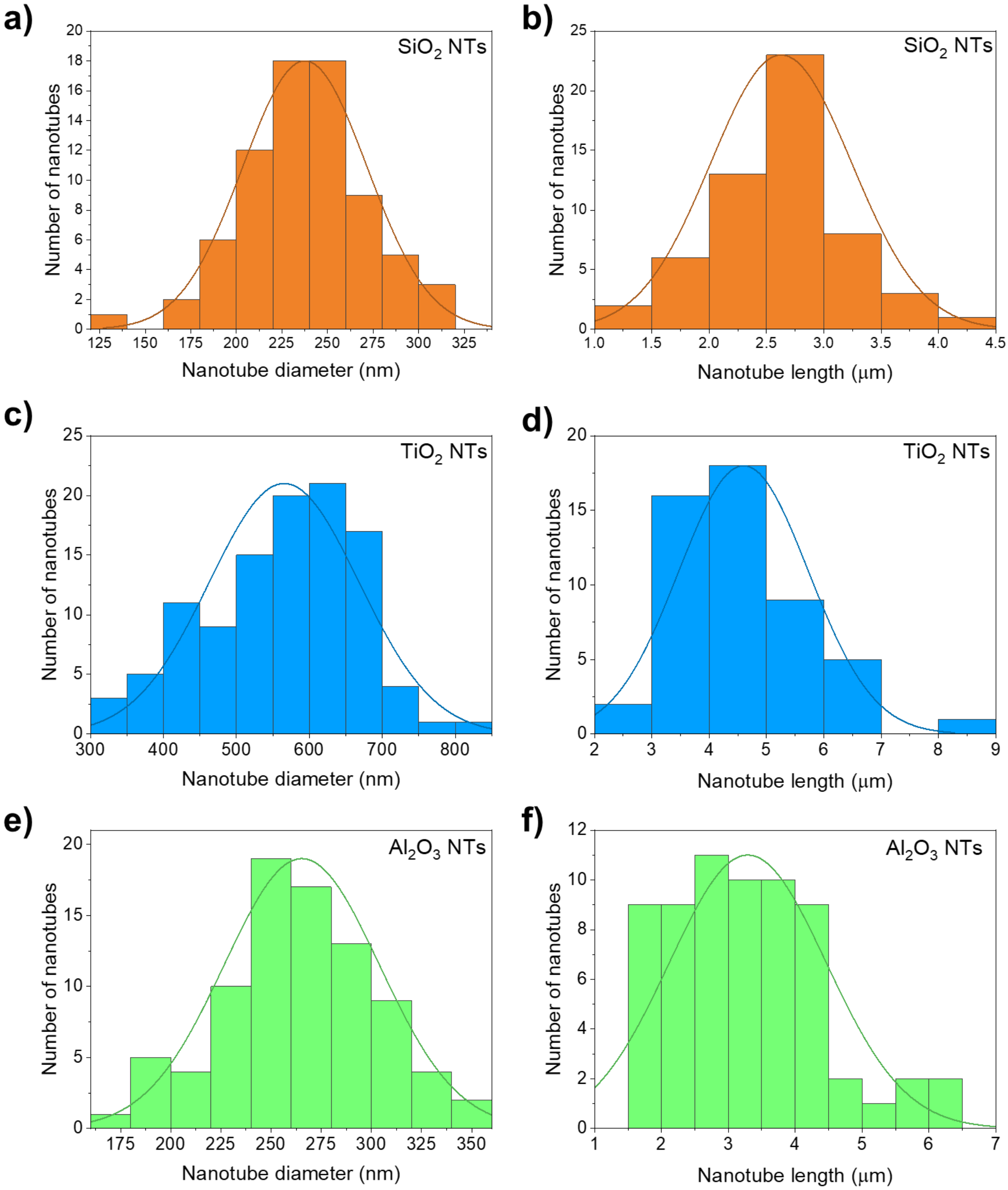

**Figure S3**. a) Cross-sectional view of graphene nanosheets over Si substrate as reference growth in standard conditions as detailed in Experimental Section. b) SEM top-view micrograph of the sample.

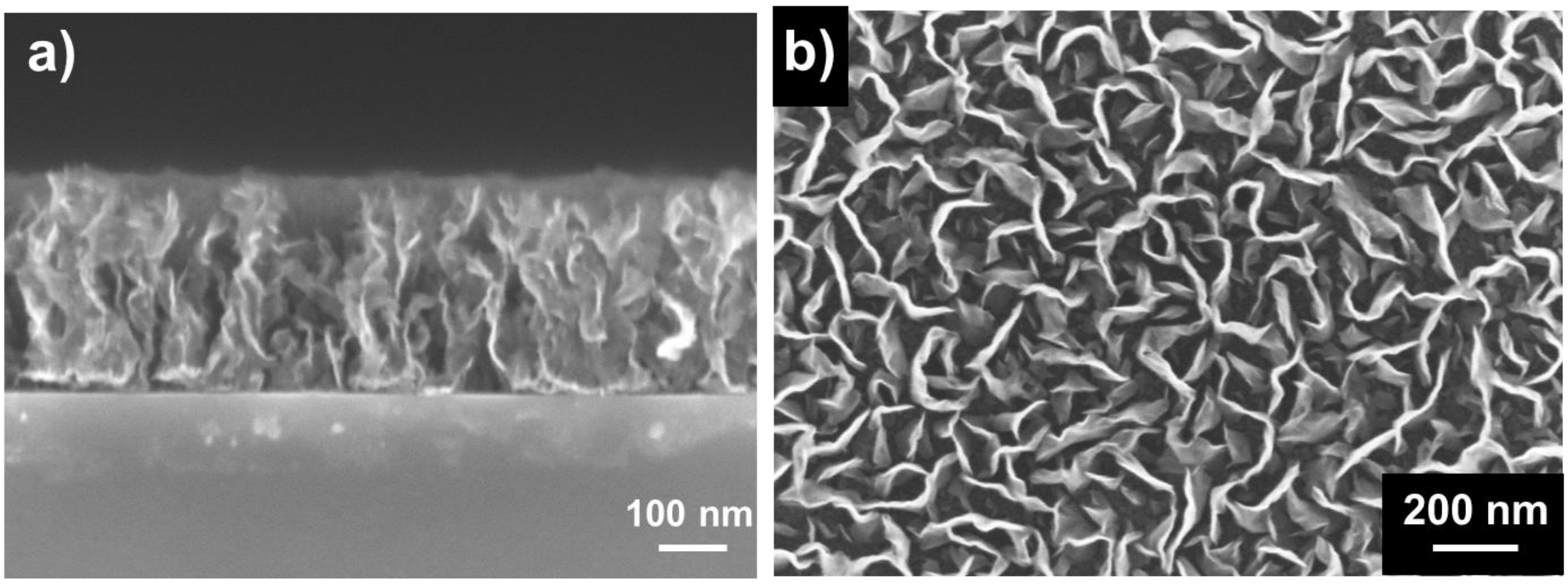

**Figure S4**. a-b) Cross-sectional view of graphene nanosheets over $SiO_2$ NTs substrate as deposited in standard conditions as detailed in Experimental Section from the bottom to the top of the nanowire ensuring a complete coverage when reduced self-shadowing is observed.

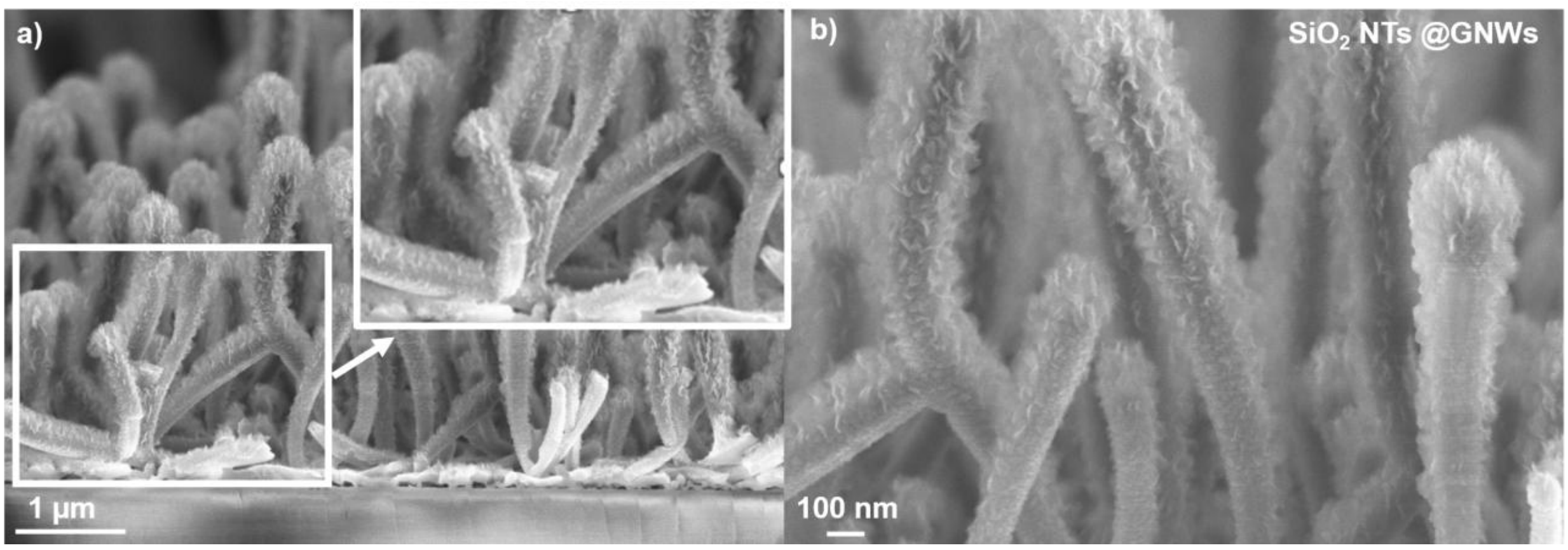

**Figure S5**. **Left**. Top view SEM micrographs of a) $SiO_2$ NTs@GNWs, b) $TiO_2$ NTs@GNWs and c) $Al_2O_3$NTs@GNWs deposited on Si wafers for the same conditions depicted in Figure 1. **Rigth.** Detailed surface of a graphene covered single nanotube.

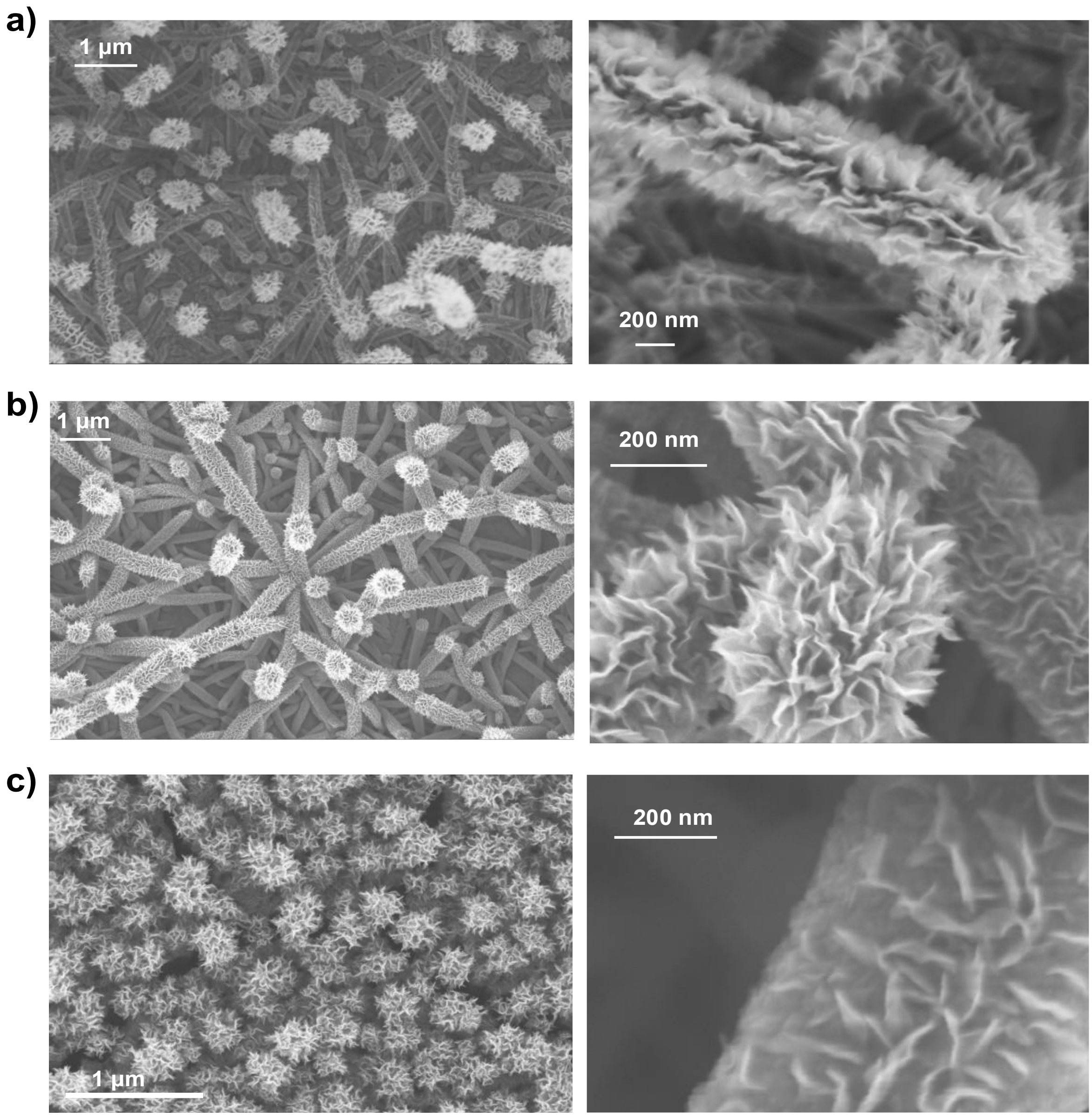

**Figure S6**. Glazing-Incident XRD analysis of $TiO_2$ NTs before and after GNWs deposition at high temperature.

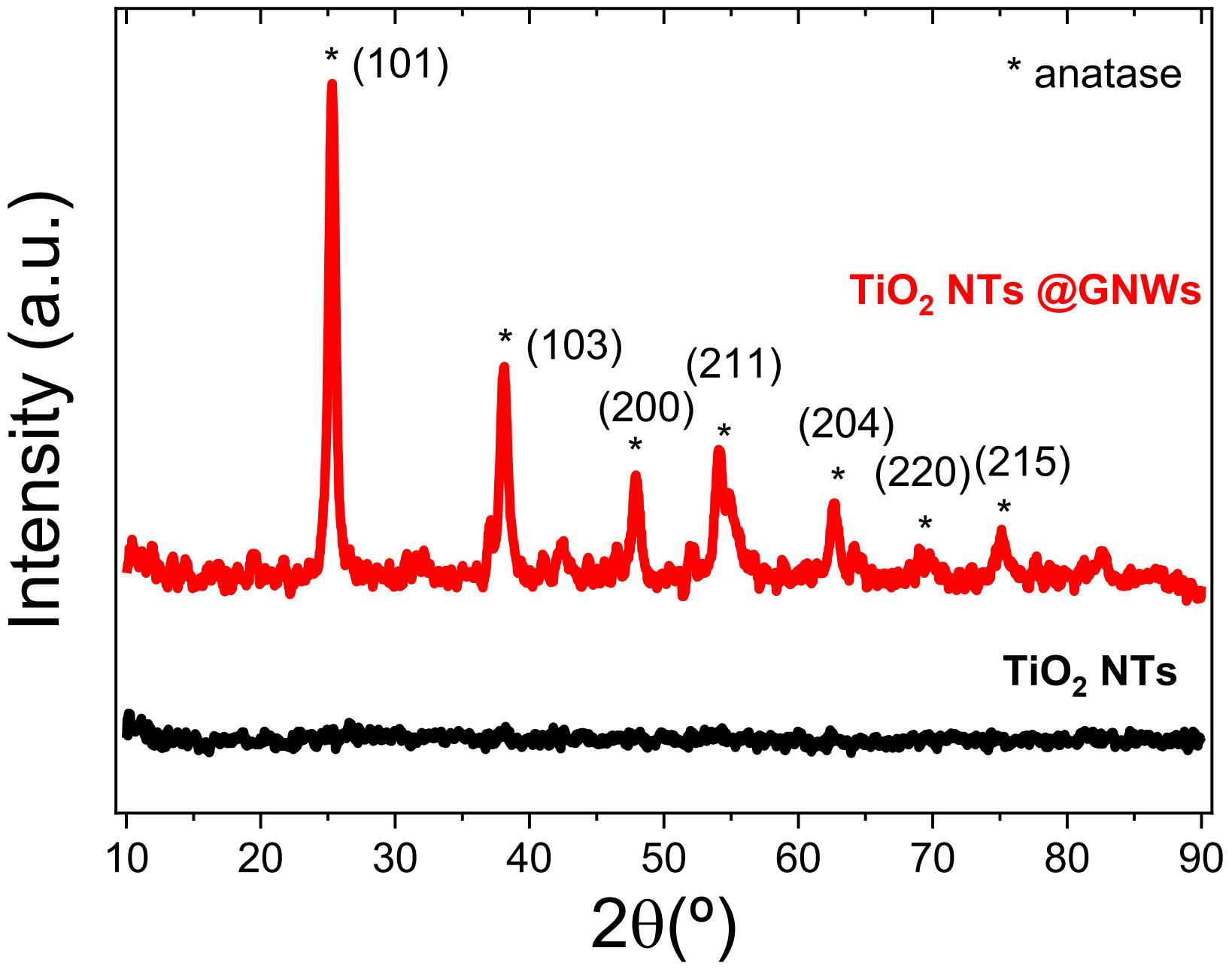

**Figure S7**. SEM images of low-temperature (below 450 ºC) graphene nanowalls over nanostructures. For each material we can see; cross-sectional as deposited $MeO_n$ NTs@GNWs, zoomed single coated nanotube and top view of a-c) $SiO_2$ NTs@GNWs, d-f) $TiO_2$ NTs@GNWs and g-i) $Al_2O_3$ NTs@GNWs.

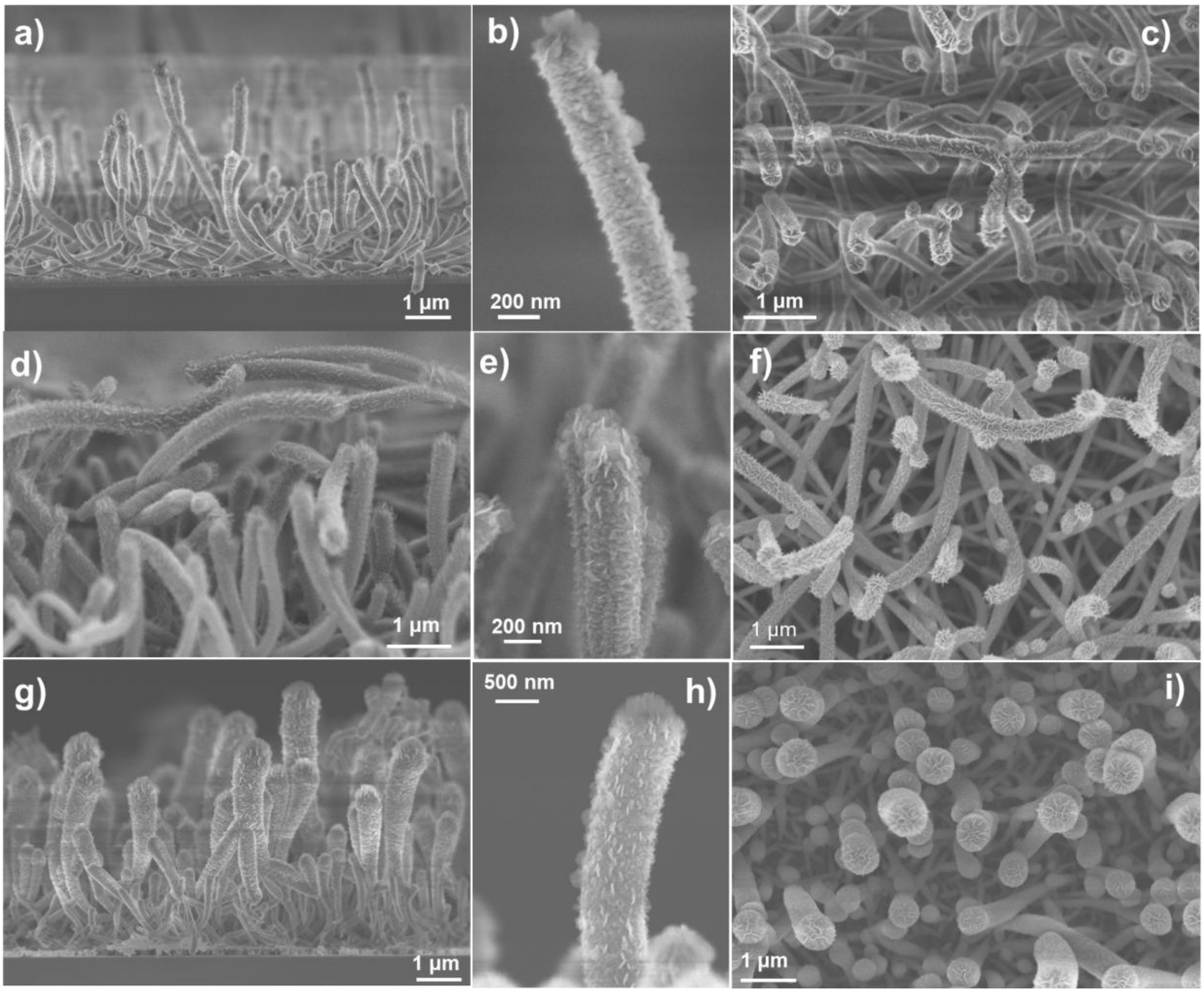

**Figure S8**. EDX profile of a single $Al_2O_3$NTs@GNWs.

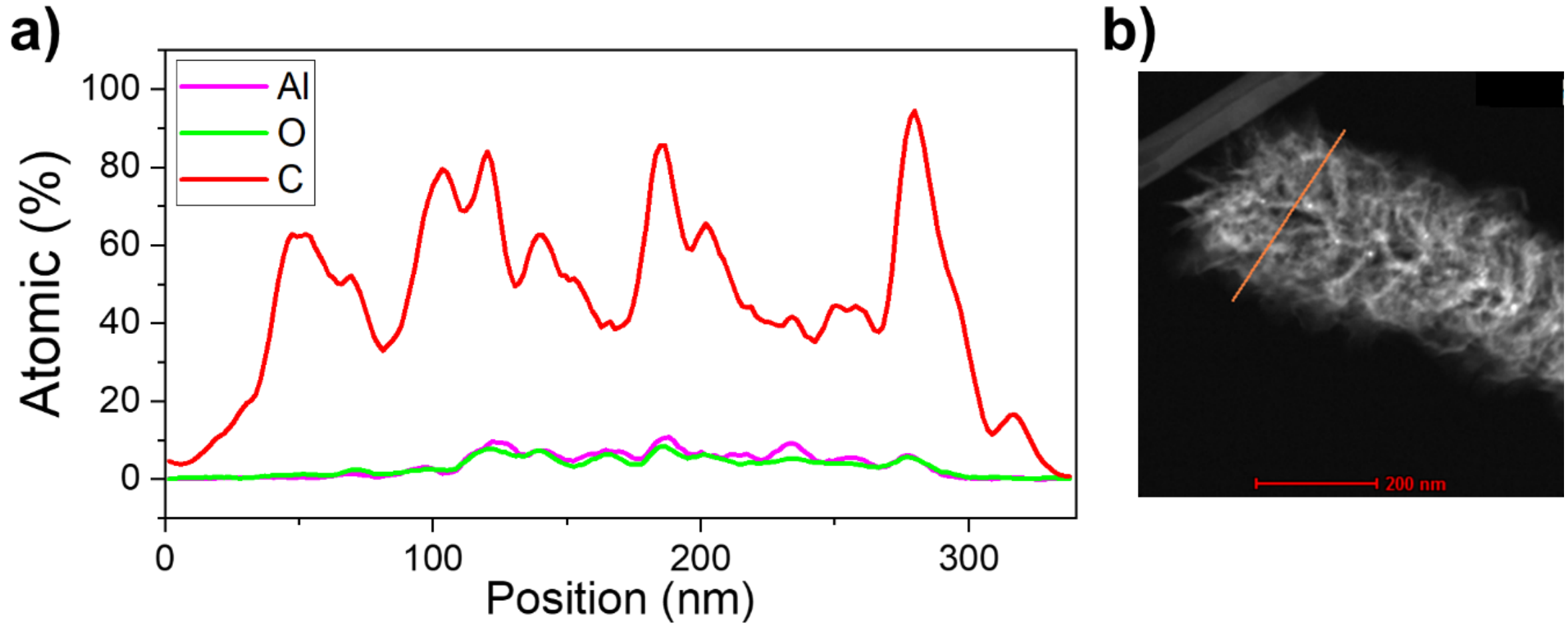

**Figure S9**. Raman Spectra of Si/GNWs growth parametrization under : a) varying RF power, b) different deposition temperatures of the substrate, c) varying space between plane parallel electrodes and d) different laser angle of incidence.

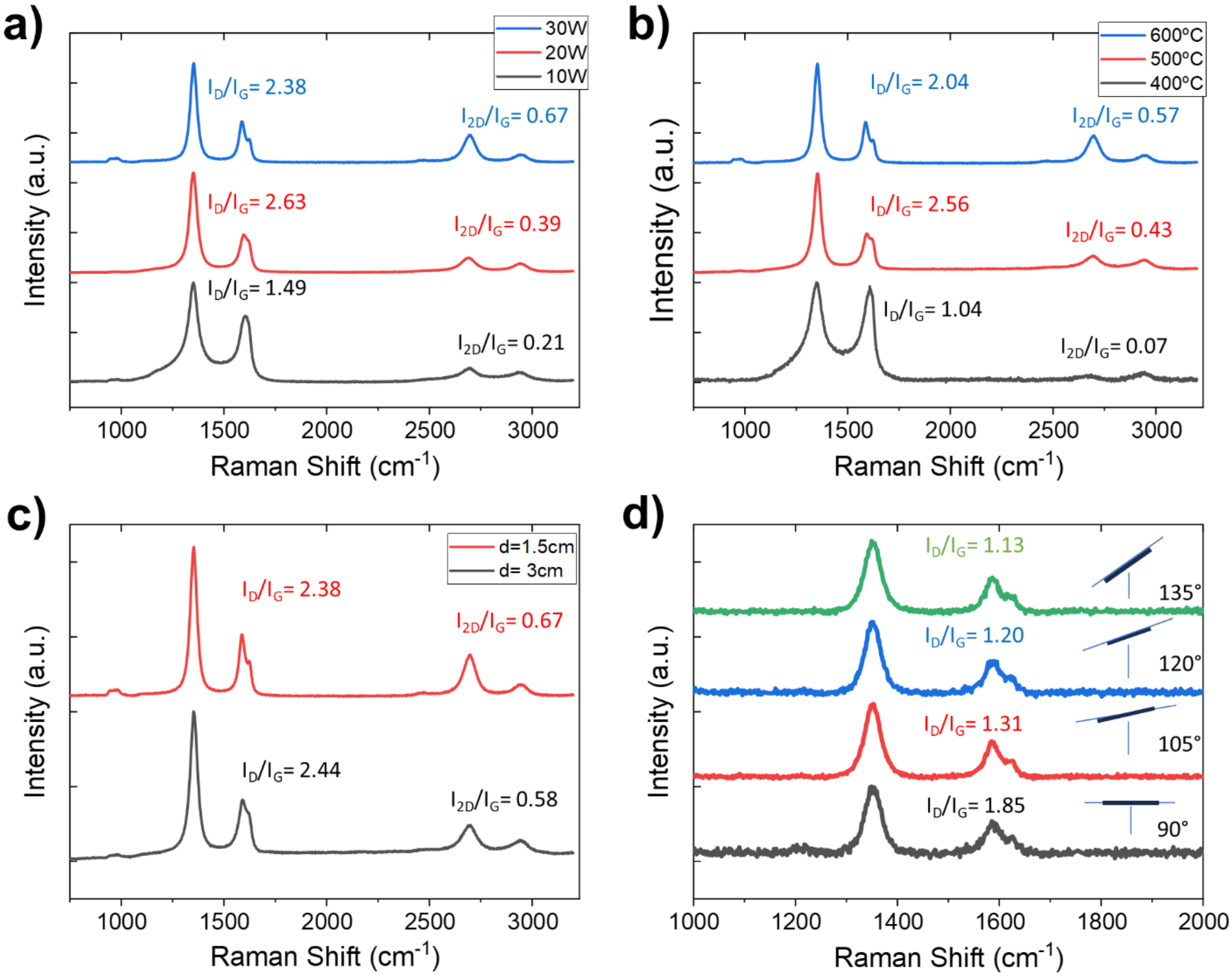

**Figure S10**. XPS survey spectra for the reference GNWs/Si (a) as well as three nanostructured supports before and after graphene deposition as $MeO_n$ NTs@GNWs for : b) $SiO_2$, c) $TiO_2$ and d) $Al_2O_3$.

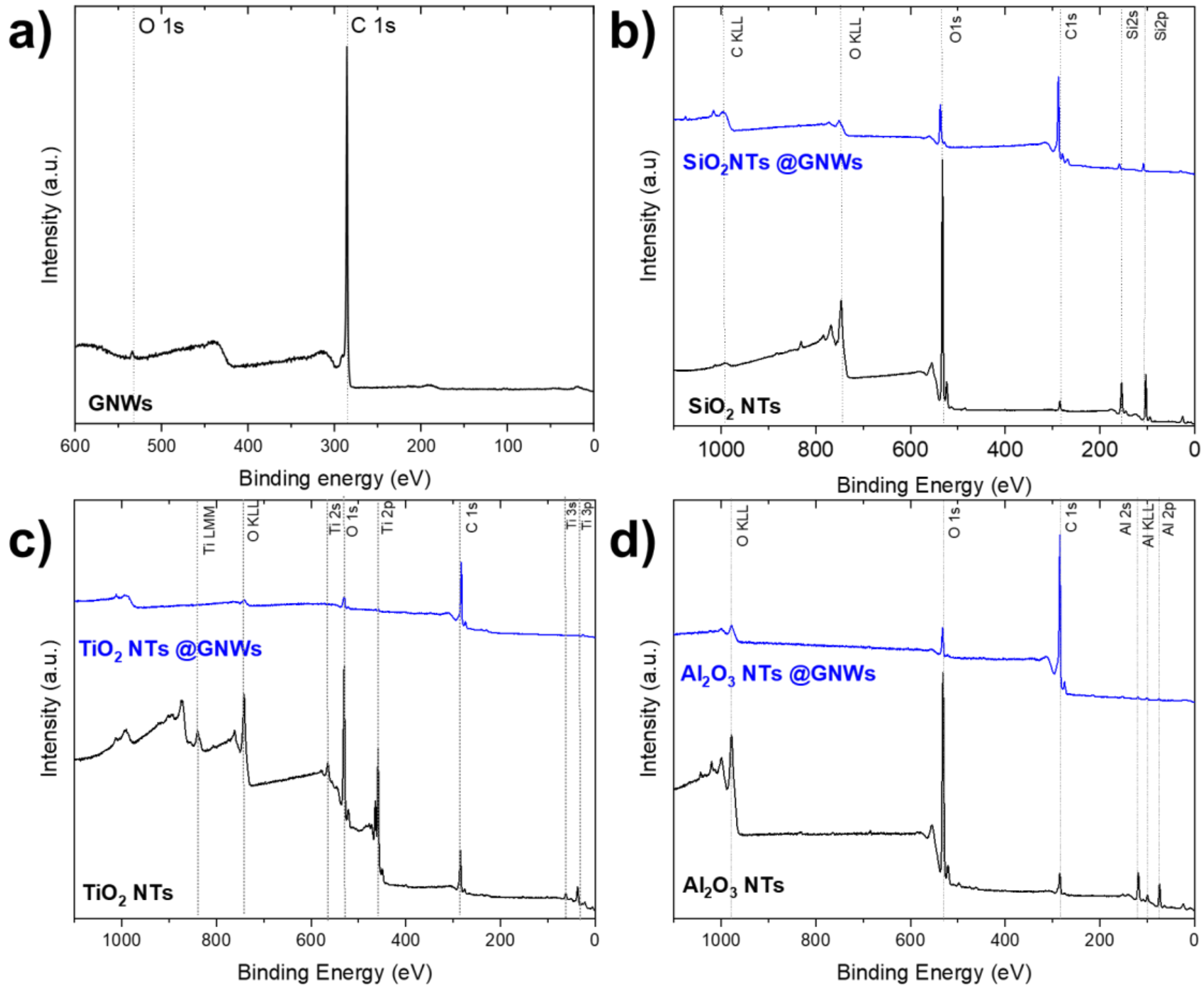

**Table S1**. Atomic concentration, metal/carbon and not-oxidized oxygen/carbon ratios by XPS analysis for the surfaces of the nanostructured supports as well as after graphene deposition. The O/C estimation of the $MeO_n$ NTs@GNWs surfaces has been obtained subtracting the contribution corresponding to the stoichiometry of the oxide composition.

| Sample | % C | % O | %Si | % Ti | % Al | Me/C | $O\text{-}O_{Me}/C$ |
|---|---|---|---|---|---|---|---|
| GNWs | 96.3 | 3.7 | - | - | - | - | - |
| $SiO_2$ NTs | 5.9 | 60.8 | 33.3 | - | - | 5.64 | - |
| $SiO_2$ NTs@GNWs | 79.6 | 19.8 | 0.6 | - | - | 0.01 | 0.234 |
| $TiO_2$ NTs | 34.7 | 46.9 | - | 18.4 | - | 0.53 | 0.291 |
| $TiO_2$ NTs@GNWs | 87.8 | 8.5 | - | 3.6 | - | 0.04 | 0.015 |
| $Al_2O_3$ NTs | 17.3 | 55.7 | - | - | 26.9 | 1.55 | 0.887 |
| $Al_2O_3$ NTs@GNWs | 91.3 | 7.2 | - | - | 1.5 | 0.02 | 0.054 |

**Figure S11.** Percentage composition of carbon functional groups resolved through the C1s XPS photopeak fitting in $MeO_n$ NTs @GNWs samples and Si/GNWs reference.

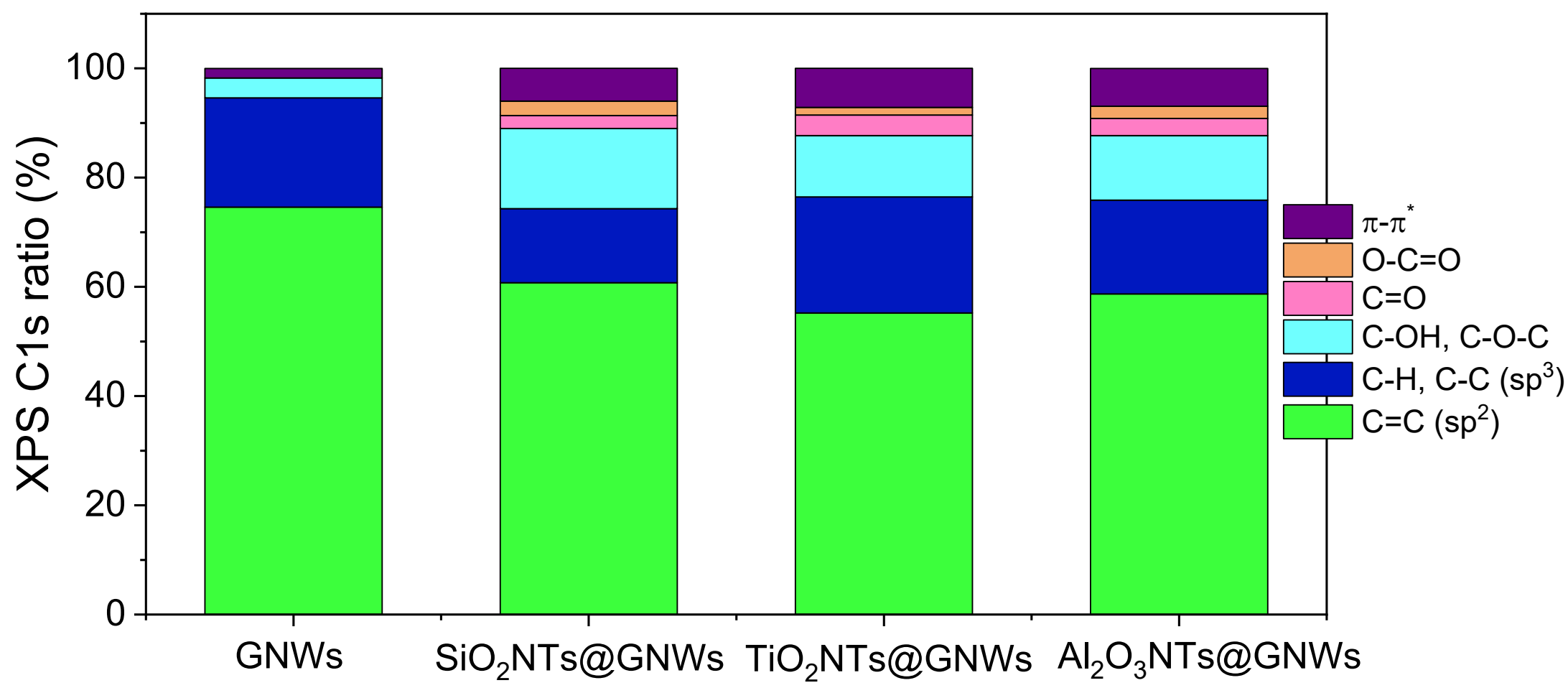

**Figure S12.** Corresponding angular dependence of the π*-resonance intensity from the NEXAFS C K edge spectra for the graphene nanowalls deposited on the nanostructured supports compared with a flat reference

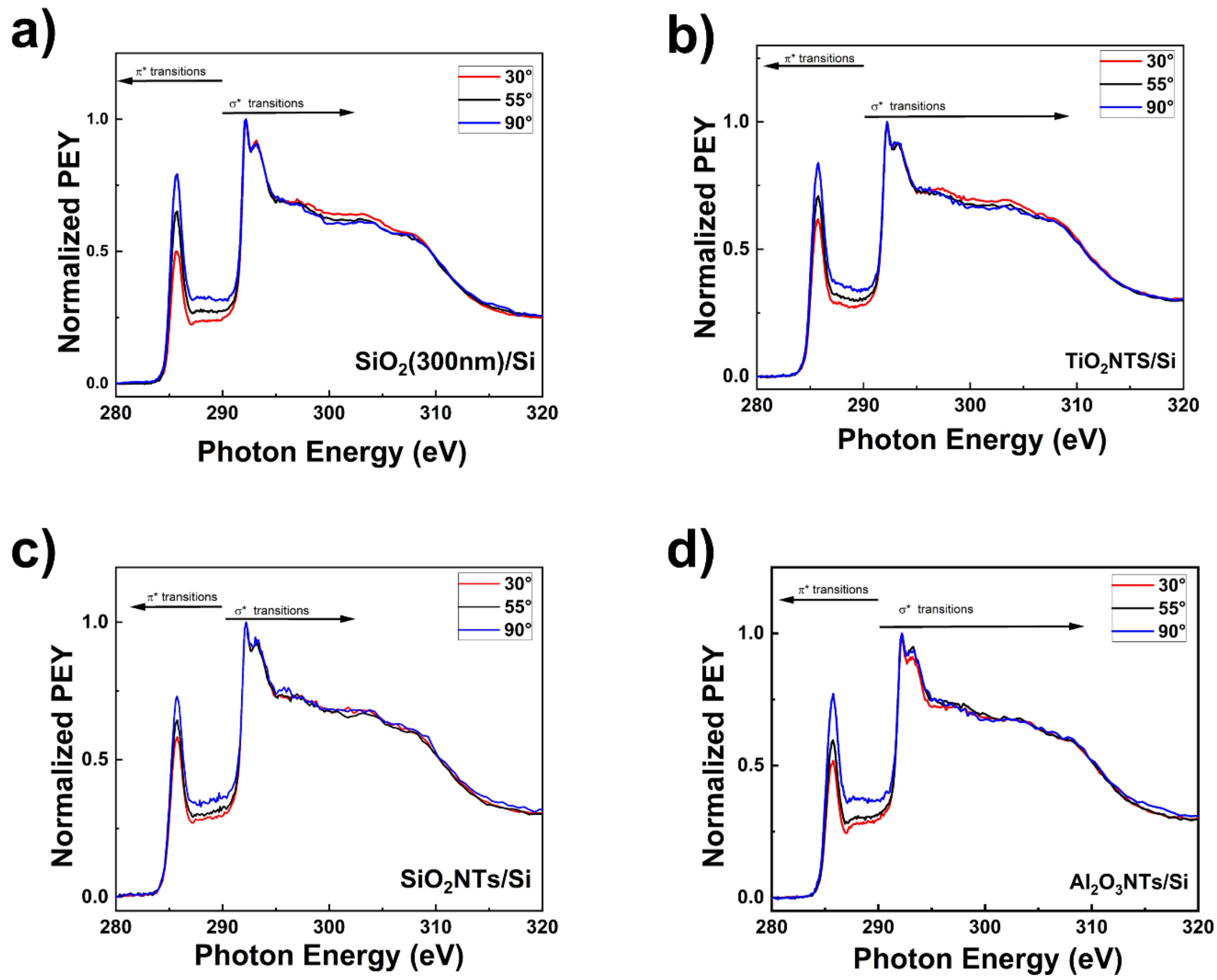

**Table S2**. Water contact angle (WCA) for two different milliQ droplets volume (2 µL and 5 µL), Contact Angle Hysteresis (CAH) for 6 µL and rolling-off angle (10 µL).

| **Sample** | **WCA (2 µL) (º)** | **WCA (5 µL) (º)** | **CAH (6 µL) (º)** | **RoA (10 µL) (º)** |
|---|---|---|---|---|
| **GNWs** | 135 | 135 | 20 | 90 |
| **$SiO_2$ NTs** | <10 | <10 | - | - |
| **$SiO_2$ NTs @GNWs** | >175 | >175 | <10 | <5 |
| **$TiO_2$ NTs** | >175 | >175 | <10 | 5 |
| **$TiO_2$ NTs @GNWs** | >175 | >175 | <10 | <5 |
| **$Al_2O_3$NTs** | >175 | >175 | <10 | <5 |
| **$Al_2O_3$ NTs @GNWs** | >175 | >175 | <10 | <5 |

**Figure S13**. a) Surface's droplet (black) and volume (red) growth rate of condensed micro-water droplets over $TiO_2$ NTs @GNWs observed in ESEM (at -1ºC, 800-810 Pa and 100% RH) from sequential images presented in b) by ImageJ software measuring average droplet diameters. Volume per water droplet has been estimated supposing a spherical geometry.

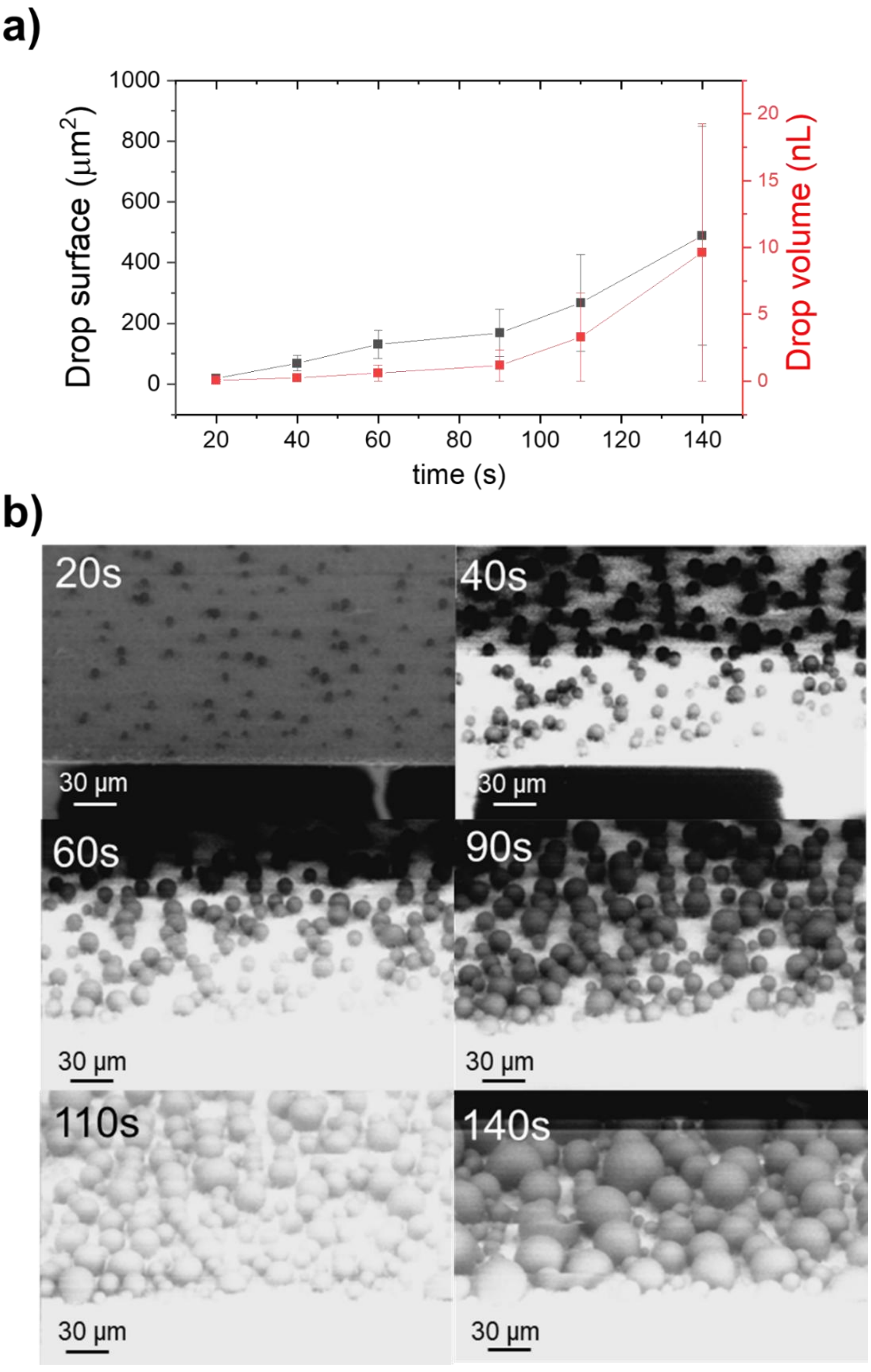